\documentclass[useAMS,usenatbib,usegraphicx]{mn2e}
\usepackage{lscape}

\newcommand{\xte}{{\textit{RXTE}}}
\newcommand{\sax}{{\textit{Beppo\-SAX}}}
\newcommand{\gro}{{\textit{CGRO}}}

\title[The synchrotron boiler]{The synchrotron boiler and the spectral states of black~hole binaries}
\author[J. Malzac and R. Belmont]{Julien Malzac\thanks{E-mail:malzac@cesr.fr} and Renaud Belmont\\
 CESR (Centre d'Etude Spatiale des Rayonnements),  Universit\'e de Toulouse [UPS], CNRS [UMR 5187],\\
9 avenue du Colonel Roche, BP 44346, 31028 Toulouse Cedex 4, France}
\begin{document}

\date{Accepted 2008 October 24,  Received 2008 October 22; in original form 2008 June 25}

\pagerange{\pageref{firstpage}--\pageref{lastpage}} \pubyear{2008}

\maketitle

\label{firstpage}

\begin{abstract}
We discuss the origin of the very different hard X-ray spectral shapes observed in the Low Hard State (LHS) and High Soft State (HSS) of accreting black holes. We study the effects of synchrotron self-absorption on the Comptonising electron distribution in the magnetised  corona of accreting black holes. 
 We solve the kinetic equations assuming that power is supplied to the coronal electrons through Coulomb collisions with a population of hot protons and/or through the injection of non-thermal energetic electrons by some unspecified acceleration process. We compute numerically the steady state particle distributions and escaping photon spectra. { These numerical simulations  confirm that synchrotron self-absorption, together with  $e$-$e$ Coulomb collisions constitute an efficient thermalising mechanism for the coronal electrons.}
When compared to the data they allow us to constrain the magnetic field and temperature of the hot protons in the corona  { independently of any dynamical accretion flow model or geometry}.
A preliminary comparison with the LHS spectrum of Cygnus X-1 indicates a magnetic field below equipartition with radiation,  suggesting  that the corona is not powered through magnetic field dissipation (as assumed in most accretion disc corona models). However, in the LHS of Cygnus X-1 and other sources, our results also point toward proton temperatures  lower than $3 \times 10^{10}$ K, i.e.  substantially lower than typical temperatures predicted by the ADAF-like models. 
In contrast, in the HSS both the proton temperature and magnetic field could be much higher. 
We also show that in both spectral states the magnetised corona could be powered essentially
through acceleration of non-thermal particles. Therefore,
contrary to current beliefs, energy dissipation processes in the corona of the HSS and that of the LHS could be of very similar nature. { The main differences between the LHS
and HSS coronal emission can then be understood as the consequence of the much stronger radiative cooling in the HSS caused by the soft thermal radiation coming from the geometrically thin accretion disc. In the LHS  the soft cooling photon flux is much weaker because the accretion disc is either truncated at large distances from the black hole, or much colder than in the HSS.}

\end{abstract}


\section{Introduction}

Black hole binaries are observed in two main spectral states. At luminosities exceeding a few percent of the Eddington luminosity\footnote{For hydrogen gas, the Eddington luminosity is $L_{\rm E}= 1.3 \times 10^{38} M$ erg s$^{-1}$ where $M$ is the black hole mass expressed in units of solar masses} ($L_{\rm E}$), the spectrum is generally dominated by a thermal component peaking at a few keV which is believed to be the signature of a geometrically thin optically thick accretion disc (Shakura \& Sunyaev 1973).
At higher energies the spectrum is non-thermal and usually presents a weak steep power-law component (photon index $\Gamma \sim 2.3-3.5$) extending at least to MeV energies, without any hint for a high energy cut-off. This soft power law is generally interpreted as inverse Compton up-scattering of soft photons (UV, soft X)  by an hybrid thermal/non-thermal distribution of electrons in a hot relativistic plasma (the so-called ``corona").  The electron distribution is then characterised by a temperature $\sim$ 30--50 keV, a Thomson optical depth $\sim$ 0.1-0.3 and a quasi power-law tail ($n(\gamma)\propto \gamma^{-s}$, with index $s$  in the range 3.5--4.5) which is responsible for most of the hard X-ray and $\gamma$-ray  emission (Gierli{\'n}ski et al. 1999, hereafter G99; Frontera et al. 2001a; McConnell et al. 2002, hereafter MC02; Del Santo et al. 2008).
Since in this state the source is bright in soft X-rays and soft in hard X-rays it is called the High Soft State (hereafter HSS).

At lower luminosities (L$<$ 0.01 $L_{\rm E}$), the appearance of the accretion flow is very different: the X-ray spectrum can be modelled as a hard power-law $\Gamma \sim 1.5-1.9$
  with a cut-off at $\sim100$ keV. The $\nu F_{\nu}$ spectrum then peaks at around a hundred keV. 
 Since  the soft X-ray luminosity is faint and the spectrum is hard, this state is called the Low Hard State (hereafter LHS). LHS spectra are generally very well fitted by  thermal Comptonisation, i.e. multiple Compton up-scattering of soft  photons by a Maxwellian distribution of  electrons (see Sunyaev \& Titarchuk 1980) in a hot ($kT_{\rm e}\sim$ 50--100 keV)  plasma of Thomson optical depth $\tau_{\rm T}$ of order unity. 
Although a weak non-thermal component is clearly detected at MeV energies in the LHS of Cygnus X-1 (MC02) and GX~339--4 (Wardzi{\'n}ski et al. 2002; Joinet et al. 2007) most of the hard X-ray luminosity emerges in the form of Comptonisation by a thermal electron distribution.  Spectral fits with hybrid thermal/non-thermal Comptonisation models suggest the slope of the non-thermal tail  in the electron distribution is steeper ($s>5$)  than in the HSS (see MC02). 

In brief, both spectral states are well represented by Comptonisation by an hybrid electron distribution. In the LHS the temperature and optical depth of the thermal electrons are higher, and the slope of the non-thermal tail seem steeper than in the HSS.  Consequently, the X-ray emission is dominated by thermal Comptonisation in the LHS and by non-thermal Comptonisation in the HSS.

The different spectral states are usually understood in terms of changes in the geometry of the accretion flow.   According to a popular scenario (see e.g. Done, Gierli\'nski and Kubota 2007, hereafter DGK07), in the HSS a geometrically thin accretion disc extends down to the last stable orbit and is responsible for the dominant thermal emission. This disc is the source of soft seed photons for Comptonisation in small active coronal  regions located above and below the disc.  The magnetic field lines rise above the accretion disc through magnetic buoyancy, transporting a significant fraction of the accretion power into the corona where it is then dissipated through magnetic reconnection (Galeev, Rosner, 
\& Vaiana, 1979).  This leads to particle acceleration in the corona.  A population of  high energy electrons is formed which then cools down by up-scattering the soft photons emitted by the disc. This produces the high energy non-thermal emission (see e.g. G99; Zdziarski et al. 2002).

In the LHS, the standard geometrically thin disc does not extend to the last stable orbit. Instead, the weakness of the thermal features suggests that it is truncated at distances ranging from a few tens to a few thousand gravitational radii from the black hole (typically  1000--10000 km). In its inner parts, the accretion flow takes the form of a hot geometrically thick, optically thin disc (Esin, McClintock and Narayan 1997; Poutanen, Krolik and Ryde 1997). In the standard hot flow solutions (see e.g. Shapiro, Lightman and Eardley, 1976;  Ichimaru 1977; Narayan and Yi 1994; Abramowicz et al 1996,  Blandford and Begelman 1999), the gravitational energy is converted through the process of viscous dissipation into the thermal energy of ions.

The ions only cool by heating up electrons through Coulomb collisions. This process being rather inefficient it leads typically to high proton temperatures ($T_{\rm i}\sim 10^{12}$ K). Such electron heating naturally forms nearly Maxwellian distributions of electrons (Nayakshin \& Melia 1998).
These electrons can then radiate their energy by Comptonising the soft photons coming from the external geometrically thin disc, as well as IR-optical photons internally generated through self-absorbed synchrotron radiation. The balance between heating and cooling determines the electron temperature which turns out to be close to 10$^9$ K, as required to fit the LHS spectra (but see Yuan \& Zdziarski 2004 and Section~\ref{sec:coulombheating}).
These two-temperature accretion flows are dominated by advection over a wide range of the parameter space, they are usually referred to as ADAF (Advection Dominated Accretion Flows).  

However, Coulomb heating is the dominant electron heating process only if the magnetic field is strongly sub-equipartition (ratios of gas to magnetic pressure $>10$). For larger magnetic fields, turbulent stresses may heat the electrons directly in the same way as they heat the protons. For approximately equipartition magnetic field the turbulence primarily heats the electrons (Quataert and Gruzinov 1999).   For even larger magnetic fields reconnection may be the dominant channel of energy dissipation into electron heating (Binovatyi-Kogan and Lovelace 1997). From an observational perspective, recent spectral modelling of the hard X-ray spectra of Sagitarius A* and black holes binaries in the LHS suggest that  a significant fraction ($\simeq$50 percent) of the energy is dissipated directly through electron heating (Yuan et al. 2003; Yuan et al. 2005; 2007) possibly through MHD turbulence, reconnection, and weak shocks.  Such processes are likely to generate a non-thermal  electron distribution function at high energy (see e.g. Li \& Miller 1997). If this is the case and if heating processes are essentially non-thermal in both spectral states, we have to understand the cause of the observed difference  in the distribution of the Comptonising electrons (i.e. thermal LHS and non-thermal in the HSS).

The main alternative to the ADAF-like models for the LHS relies on electron heating by
magnetic reconnection above the standard thin disk (Liang \& Price 1977; Galeev et al 1979) as discussed earlier for  the HSS.  
Since the process of magnetic reconnection remains poorly understood, this
model has not yet been elaborated in detail, although numerical simulations have shed
some light on this problem (Miller \& Stone 2000; Hirose, Krolik \& Stone 2006). However the free parameters of this accretion disc corona model  (see Haardt, Maraschi \& Ghisellini 1993) are very constrained  by the observations (see Malzac 2007 for a discussion). This model successfully reproduces LHS spectra by  postulating the presence of outflowing magnetic active region  above a cold thin disc extending close to the last stable orbit (Beloborodov 1999; Malzac, Beloborodov \& Poutanen 2001).  In this framework, the thin disc is never truncated, simply colder in the LHS (see Malzac 2007). This dynamic corona is assumed to be powered through the same magnetic processes as the accretion disc corona of the HSS.  However  the cause for the different electron distributions was never discussed so far, which is precisely the aim of this paper.

The main idea we investigate here is that in presence of a magnetic field, the electron distribution can appear thermal even when acceleration mechanisms would produce non thermal distributions.  Synchrotron self-absorption indeed represents a very fast and efficient thermalising mechanism. The importance of this effect for the emission of compact sources was first pointed out by Ghisellini, Guilbert and Svensson (1988). 
These authors and then Ghisellini, Haardt and Svensson (1998) considered the case of high energy electrons continuously injected in a magnetised region and then cooling  down through synchrotron and inverse Compton emission. Solving the kinetic equations, they showed that due to the very fast emission and absorption of synchrotron photons, the electron distribution relaxes toward a Maxwellian distribution in a few light crossing times. They named this effect the synchrotron boiler.  The steady state distribution then consists in a Maxwellian with a quasi-powerlaw high energy tail (formed by the cooling electrons). { In the context of ADAF models,  Mahadevan and Quataert (1997) showed  that the synchrotron boiler  together with $e$-$e$ Coulomb collisions would efficiently thermalise the electron distribution of the hot accretion flow}.  

Although these studies demonstrated the correct qualitative behaviour, they were based on a very simplified treatment, in which the electron and photon kinetic equations are not solved self-consistently and the Compton process is modelled in the Thomson approximation.  Although this treatment provides important insights, it is not realistic enough for a quantitative comparison with data. 
Recently, Vurm and Poutanen (2008), presented the results of a computer code which overcomes these problems and confirms the earlier results.
 Their code solves simultaneously the kinetic equations for photons and electrons using the full Klein-Nishima cross section for Compton scattering.  Their simulations however do not  account for other radiation processes such as pair production, or other, possibly important, thermalising mechanisms such as Coulomb interactions { between electrons}. 
Several other computer codes such as {\sc eqpair} (Coppi 1992, 1999) or the Fokker-Planck code by  Nayakshin \& Melia (1998) take into account these processes,  but not  the thermalising effects of the magnetic field.  Moreover,  {\sc eqpair}  postulates \emph{a priori} a low energy thermal electron distribution, which is not ideal when studying thermalisation processes.

The first full treatment of the problem (including  all the relevant processes) under the 1-zone, homogeneous, isotropic approximation, was performed only recently with a code presented in 
Belmont, Malzac \& Marcowith (2008a, 2008b).  In the present paper we use this code to investigate in more details the synchrotron boiler effects  in physical situations relevant to the corona of accreting black holes. We also present a first direct comparison of our simulations with the data. 

 In Section~\ref{sec:model} we present the assumptions of our model  and derive useful analytical estimates for the temperature of the hot protons and the equi-partition magnetic field as a function 
 of compactness. In Section~\ref{sec:simres} we explore the parameter space. Assuming that electrons are accelerated according to a power-law distribution, we show that, for reasonable parameters, the steady state spectrum is very similar to that observed in the LHS of black hole binaries. Then we show that the change in the thermal vs non-thermal Compton emission  between the LHS and HSS can be understood as the effect of the strong soft  photons flux cooling down the thermal part of the electron distribution to temperatures at which its Compton emissivity is negligible and leaving a bare non-thermal spectrum.  We also discuss the effects of Coulomb heating by hot protons.
 Finally, in Section~\ref{sec:cygx} we attempt to model observed LHS and HSS spectra of Cygnus X-1. This allows us to put some constraints on the magnetic field and proton temperature of the coronal plasma in both states. This comparison also confirms that both spectral states can be modelled with power-law  electron acceleration as unique heating mechanism. These results are then summarised and discussed in Section~\ref{sec:discussion}.

\section{Model}\label{sec:model}

\subsection{Framework}

We use the code of Belmont et al. (2008b). This code solves the kinetic equations for photons, electrons and positrons  accounting for Compton scattering (using the full Klein-Nishima cross section), electron-positron pair production and annihilation, Coulomb interactions { (electron-electron and electron-proton)}, synchrotron emission and absorption and $e$-$p$ bremsstrahlung. Radiative transfer is dealt using a usual escape probability formalism. 
This code was critically tested and found consistent with the results of the {\sc eqpair} code (Coppi 1992) and the Fokker-Planck code of Nayakshin and Melia (1998) which however do not account for synchrotron.
 
 { Our  code is fully time-dependent. 
 An example of time-dependent simulation of the synchrotron boiler effect is presented in Belmont et al. (2008b). For compactness parameters of order of unity, or larger, the system relaxes toward equilibrium very quickly (typically a few  light-crossing times).
This relaxation time is smaller than, (or at most comparable to) the dynamical time-scale of the accretion disc at all radii\footnote{ The light crossing time $t_{\rm light}$ of an emitting region of size $R$, is  always much shorter than the local dynamical time-scale $t_{\rm dyn}$ at a distance $R$ from a black hole. For a black hole of gravitational radius $R_{\rm G}$,  $t_{\rm dyn}/t_{\rm light}=2\pi \sqrt{R/R_{\rm G}}$}. Moreover the  observation shows that the X-ray variability of black hole binaries and active galactic nuclei is dominated by fluctuations on time-scales that are much longer than the light-crossing time of the inner disc region which is supposed to produce those X-rays (see e.g. DGK07). 
Therefore, for most applications assuming steady state should be a good approximation. 
In this paper we limit our study to steady-state particle distributions and photon spectra.}

We consider a sphere with radius $R$  of fully ionised proton-electron magnetised plasma in steady state. The Thomson optical depth of the sphere is $\tau_{\rm T}=\tau_{\rm i}+\tau_{\rm s}$, where $\tau_{\rm i}=n_i\sigma_{\rm T}R$ is the optical depth of ionisation electrons (associated with protons of density $n_i$) and $\tau_s=2n_{\rm e^{+}}\sigma_{\rm T}R$ is the optical depth of electrons and positrons due to pair production ($n_{\rm e^{+}}$ is the positron number density). $\sigma_{\rm T}$ is the Thomson cross section.
 The radiated power is quantified through the usual compactness parameter:
 \begin{equation}
 l=\frac{L\sigma_{\rm T}}{R m_{\rm e} c^3}= 2.7\frac{L}{10^{37} \rm{erg}}\frac{10^8 \rm{cm}}{R}=7.7 \frac{L}{10^{-2} L_{\rm E}}\frac{30 R_{\rm G}}{R},
 \label{eq:l}
 \end{equation}
where $L$ is the luminosity of the Comptonising cloud,  $m_{\rm e}$ the electron rest mass and $c$ the speed of light. We introduced the Eddington luminosity $L_{\rm E}=1.3 \times 10^{8} M/M_{\odot}$ erg s$^{-1}$, and the gravitational radius $R_{\rm G}=GM/c^2$, where $G$ is the gravitational constant and $M$ the mass of the black hole.

\subsection{Power supply}

We will consider three possible channels for the energy injection in our coupled electron-photon system: 
\begin{enumerate}
\item Non-thermal electron acceleration  with a compactness $l_{\rm nth}$.
We model the acceleration process by assuming electrons are continuously injected with a power-law distribution of index $\Gamma_{\rm inj}$  (i.e. $n(\gamma)\propto \gamma^{-\Gamma_{\rm inj}}$), between Lorentz factors $\gamma_{\rm min}$  and $\gamma_{\rm max}$. Observations of Cygnus X-1 in the soft state indicate $\Gamma_{\rm inj}$  in the range 2.0--3.5 (G99). { In most of our simulations,} we set $\gamma_{\rm min}=1$, $\gamma_{\rm max}=1000$. { As will be shown in section~\ref{sec:nonthe},} as long as $\gamma_{\rm min}<2$ and $\gamma_{\rm max}\gg1$, and $\Gamma_{\rm inj}>2$ the exact values have little effects on the result. 
 In this paper, we assume that the particles are reaccelerated rather than injected, and therefore we balance the number of 'injected' non-thermal particles by removing the same number of particles from the steady state distribution. We define the non-thermal compactness as the net energy input due to the acceleration process, i.e:
\begin{equation}
l_{\rm nth}=[\langle\gamma\rangle_{\rm i}-\langle\gamma\rangle]\dot{n} \sigma_{\rm T}/(Rc)\label{eq:lnth}
\end{equation}
where $\dot{n}$ is the total number of injected/accelerated particles per unit time,  $\langle\gamma\rangle$  is the average Lorentz factor of electrons in the hot plasma,  and $\langle\gamma\rangle_{\rm i}$ is the average Lorentz factor at which they are accelerated/injected  (which depends on $\gamma_{\rm min}$, $\gamma_{\rm max}$ and $\Gamma_{\rm inj}$). This definition of $l_{\rm nth}$ corresponds to that of the version of {\sc eqpair} used by MC02 and other authors to fit the spectra of black hole binaries. { We note that this assumes that re-acceleration occurs uniformly at all energies up to $\gamma_{\rm max}$ (while in some other acceleration models the accelerated particles are drawn from the thermal part of the distribution).}\\

\item  Coulomb heating with a compactness $l_{\rm c}$. 
Electrons are supposed to interact by Coulomb collisions with a distribution of thermal protons. When the protons have a larger temperature than the electrons, the electrons gain energy and $l_{\rm c}>0$ (see below Eq. \ref{eq:coul}). In our model, we fix $l_{\rm c}$,  then determine the proton temperature corresponding to this compactness and to the exact steady state electron distribution.
 
 In the case of a pure Maxwellian plasma, the dependence of the Coulomb compactness on electron and proton temperatures can be approximated as follows (Dermer 1986): 
\begin{equation}
l_{\rm c}=\sqrt{8\pi}\tau_{\rm T}\tau_{\rm i}\ln\Lambda(\theta_{\rm i}-\frac{m_{\rm e}}{m_{\rm i}}\theta_{\rm e})(\theta_{\rm e}+\theta_{\rm i})^{-3/2},
\label{eq:coul}
\end{equation}
where $\ln\Lambda\simeq20$ is the Coulomb logarithm and $\theta_{\rm e}=\frac{kT_{\rm e}}{m_{\rm e}c^2}$ is the electron temperature in units of the electron rest mass. Equation~\ref{eq:coul} is valid in the limit of small $\theta_{\rm i}$ and $\theta_{\rm e}$.  For typical values  $\theta_{\rm e}\simeq0.18$ and  $\theta_{\rm i}\simeq0.01$ , $l_{\rm c}\simeq 12 \tau_{\rm T}\tau_{\rm i}$. 

To the first order in $\theta_{\rm i}/\theta_{\rm e}$ equation~(\ref{eq:coul}) can be rewritten as:
\begin{equation}
\frac{T_{\rm i}}{T_{\rm e} }\simeq1+\frac{m_{\rm i}}{m_{\rm e}}\frac{g l_{\rm c} }{\sqrt{56\pi}\ln\Lambda\tau_{\rm T}\tau_{\rm i}}\simeq 1+7\frac{gl_{\rm c}}{\tau_{\rm T}\tau_{\rm i}},
\label{eq:temp}
\end{equation}
where $g(\theta_{\rm e})=\sqrt{7\theta_{\rm e}}$ ($\simeq1$ in the LHS). The proton temperatures provided by equation~(\ref{eq:temp}) agree with the values obtained from our numerical simulations  within 30 percent, even when the lepton distribution strongly departs from a Maxwellian.  The largest discrepancies occur when  the approximation of small $\theta_{\rm i}/\theta_{\rm e}$ breaks down. For the typical electrons temperatures observed in accreting black hole sources, this happens for proton temperatures $>50$ MeV. In sources with accurately measured electron temperature, Thomson optical depth, and luminosity, and for a given size of the emitting region, equation~(\ref{eq:temp}) provides an upper limit on the proton temperature. It will be  used in Sections~\ref{sec:cygx} and~\ref{sec:coulombheating} to constrain the temperature in Cyg X-1 and XTE~J1118+480. \\

\item External soft radiation coming from the geometrically thin accretion disc and entering the corona with a compactness $l_{\rm s}$, that we model as homogeneous injection of photons with a blackbody spectrum of temperature $kT_{\rm bb}$.
\end {enumerate}
Since, in this model, all the injected power ends up into radiation we have: $l=l_{\rm nth}+l_{\rm c}+l_{\rm s}$ in steady state.

We note that this set-up is very similar to the one used in the {\sc eqpair} model when fitting the data of black hole binaries (see G99).  The main difference is that we take into account the synchrotron emission and absorption processes and the associated synchrotron boiler effect. In our code the thermalisation process is treated self-consistently while in {\sc eqpair} it is assumed that the electron distribution is Maxwellian at low energies. Another difference is that we have a detailed description for Coulomb heating by hot protons. In {\sc eqpair}, electron heating is parametrised through a `thermal' compactness $l_{\rm th}$ quantifying the amount of power provided as heating to the thermal pool of cool electrons. Actually, $l_{\rm th}$ plays a very similar role as $l_{\rm c}$ in our model. There is a difference however, which is that in our model not only the thermal but also the non-thermal leptons are allowed to gain (or loose) energy through Coulomb interactions with the protons.
In the {\sc eqpair} model the results are usually parametrized as a function of the non-thermal fraction $l_{\rm nth}/l_{\rm h}$ where $l_{\rm h}$=$l_{\rm nth}$+$l_{\rm th}$  and $l_{\rm nth}$ is defined as in equation~(\ref{eq:lnth}). In our model such parametrization would be misleading since, as it will be  shown below, a large fraction of the power provided as non-thermal electron acceleration (i.e. $l_{\rm nth}$ ) actually ends up heating the lowest energy particles of the distribution via the synchrotron and Coulomb boilers. 

\subsection{Magnetic field}

\begin{figure*}
 \includegraphics[width=\textwidth]{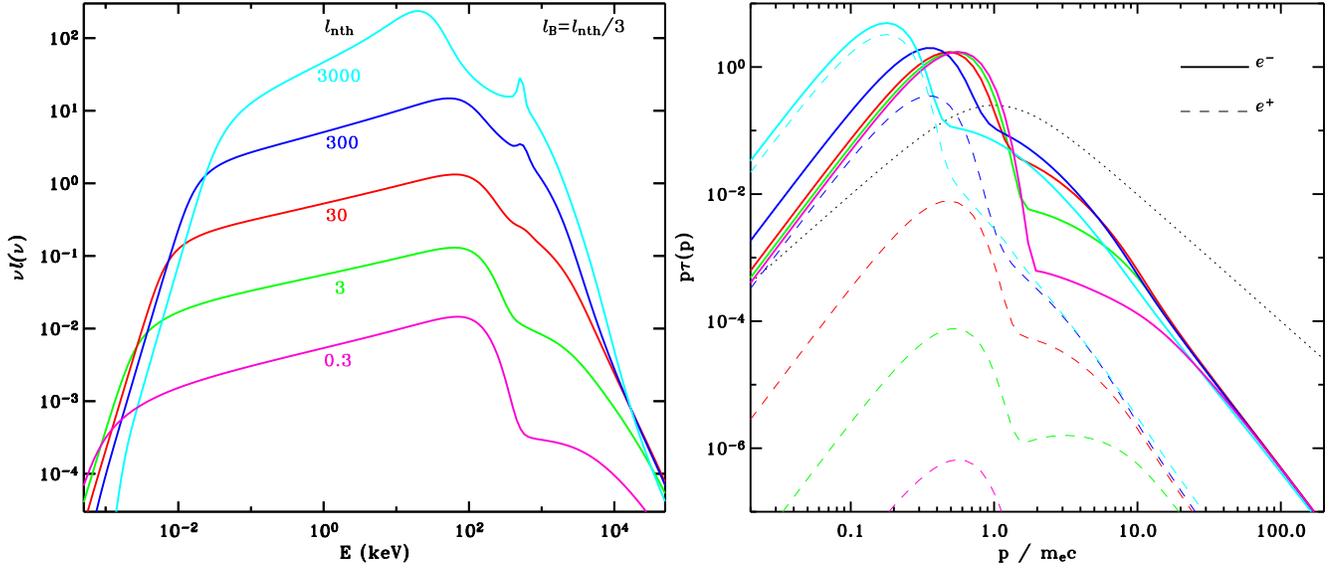}
 \caption{Synchrotron self-Compton models with pure non-thermal injection. Effects of varying $l_{\rm nth}$ at constant $l_{\rm nth}/l_{B}$. Simulated photon spectra (left panel) and particle distributions for $l_{\rm nth}$= 0.3 (pink), 3 (green), 30 (red),  300 (blue),  3000 (cyan) and $l_{\rm nth}/l_{B}=3$. The other parameters are $\Gamma_{\rm inj}=3$, $\gamma_{\rm min}=1$, $\gamma_{\rm max}=1000$, $\tau_{\rm i}=2$, $R=5\times 10^7$ cm. In the right hand side panel the full curves show the electron, while the dashed ones show the positron distributions. { The dotted curve shows the shape of the injected electron distribution.}}
 \label{fig:le}
\end{figure*}

While the total injection compactness $l$  can be evaluated using size estimates and the observed luminosity of the sources, the strength of the tangled magnetic field $B$ is the most uncertain parameter of the problem.  It is parametrized through the usual magnetic compactness:
\begin{equation}
l_{B}=\frac{\sigma_{\rm T}}{m_{\rm e} c^2}R \frac{B^2}{8\pi},
\end{equation}
or equivalently
\begin{equation}
B=5.9\times 10^5 \left[l_B\ \frac{30 R_{\rm G}}{R}\frac{20M_{\odot}}{M}\right]^{1/2} {\rm G}.
\end{equation}

The results of this paper will be discussed mostly in two different situations, namely for magnetic field in equipartition with the radiation field and with the proton energy density.

If the magnetic field is at equipartition with the radiative energy density then:
\begin{equation}
\frac{B_{\rm R}^2}{8\pi}=U_R=L t_{\rm esc}/V,
\end{equation}
where $t_{\rm esc}$ is the photon escape time { and $V$ is the volume of the sphere}. We can adopt the prescription of Lightman \& Zdziarski (1987) which is used in our code and which, in the limit of low energy photons ($h\nu<m_{\rm e}c^2$), gives  $t_{\rm esc}\simeq(1+\tau_{\rm T}/3)R/c$. 
One obtains an equipartition magnetic compactness of:
\begin{equation}
l_{B_{\rm R}}\simeq\frac{3l}{4\pi}(1+\frac{\tau_{\rm T}}{3}).
\label{eq:eqprad}
\end{equation}

Alternatively, if the magnetic field is in equipartition with the ion energy density then:
\begin{equation}
l_{B_{\rm P}}=\frac{3}{2}\tau_{\rm i}\frac{m_{\rm i}}{m_{\rm e}}\theta_{\rm i}\simeq29 \tau_{\rm i} \frac{T_{\rm i}}{T_0},
\label{eq:beq}
\end{equation}
where $m_{\rm i}/m_{\rm e}=1836$ is the proton to electron mass ratio, $\theta_{\rm i}=kT_{\rm i}/m_{\rm i} c^2$ is the proton temperature in units of the proton rest mass energy. $kT_0=10$ MeV is a typical value for the temperature in the innermost part of an advection dominated accretion flow. This gives an estimate of the equipartition magnetic field:  
\begin{equation}
B_{\rm P}=3.2 \times 10^6 G \left(\tau_{\rm i}\frac{30R_{\rm G}}{R} \frac{20M_{\odot}}{M} \frac{T_{\rm i}}{T_0} \right)^{1/2}. 
\end{equation}

It is also interesting to express the equipartition magnetic compactness as a function of the Coulomb compactness. 
For magnetic field in equipartition with the proton energy density, we can combine equation~(\ref{eq:temp}) and~(\ref{eq:beq}) in order to estimate the equipartition magnetic compactness as a function of $l_{\rm c}$:
\begin{equation}
l_{B_{\rm P}}\simeq\frac{3}{8}y\left(\frac{\tau_{\rm i}}{\tau_{\rm T}}+7 g \frac{l_{\rm c}}{\tau_{\rm T}^2}\right),
\end{equation}
where $y=4\tau_{\rm T}\theta_{\rm e}$ is the Compton parameter.

In the limit of large Coulomb compactness (or small Thomson depth),  $l_{B_{\rm P}}\simeq 3 y g l_{\rm c}/\tau_{\rm T}^2$.
For $\tau_{\rm T}=2$ and $yg$$\simeq1$,  this leads to $l_{B_{\rm P}} \simeq 0.7l_{\rm c}$ which, in absence of other sources of heating or soft photons,  is comparable  to the values obtained assuming equipartition with radiation ($l_{B_{\rm P}}=0.4l$).
In the limit of small coulomb compactness, however, the  equipartition magnetic compactness saturates at $3y/8\simeq0.4$ corresponding to equipartition in a one temperature plasma ($B_{\rm P}\simeq 5 \times 10^5$ G for typical parameters). 
\begin{figure*}
 \includegraphics[width=\textwidth]{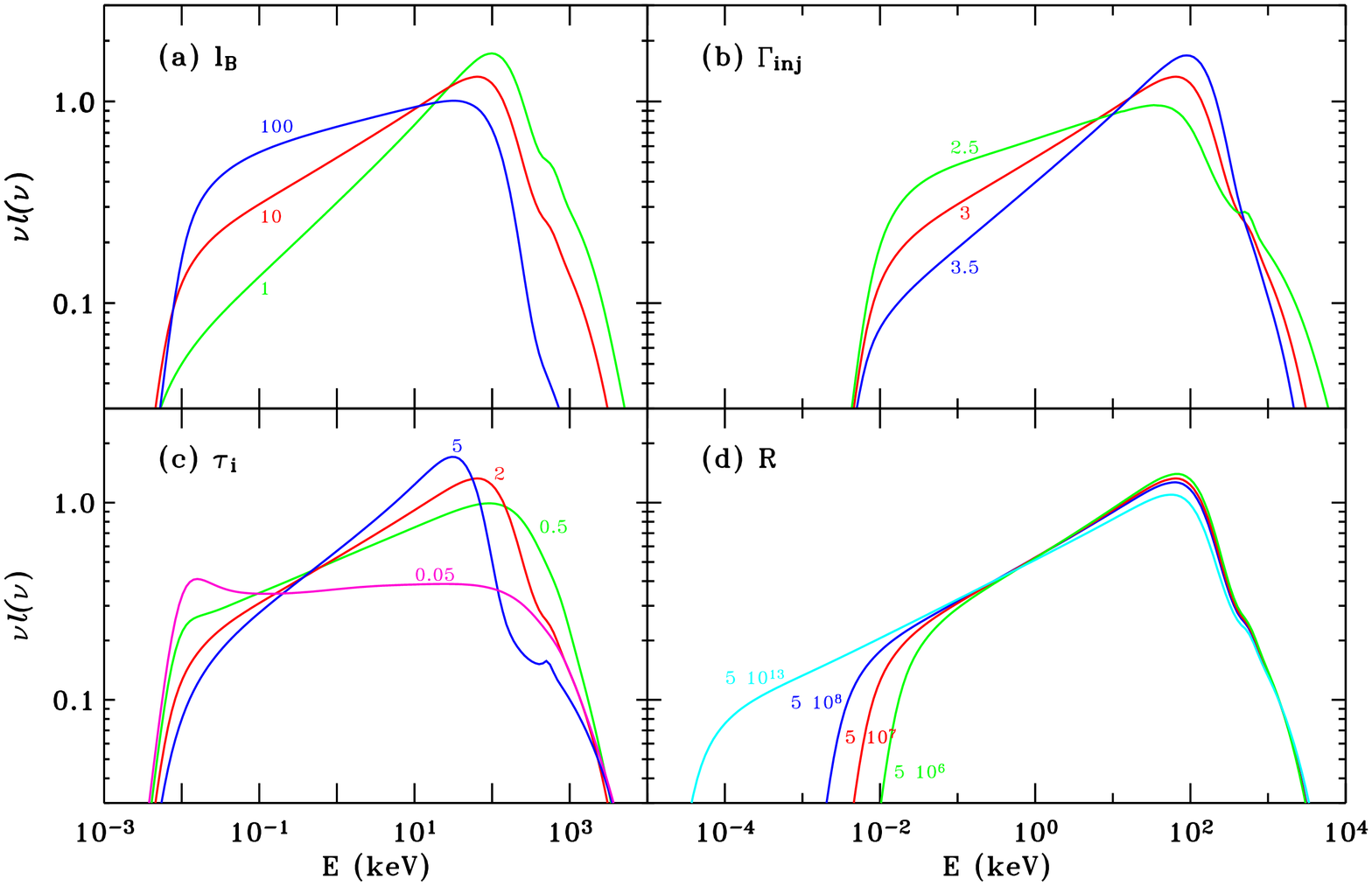}
 \caption{{ Synchrotron self-Compton models with pure non-thermal injection.} Effects of varying $l_B$ (panel a), $\Gamma_{\rm inj}$ (panel b), $\tau_{\rm i}$ (panel c) and $R$ (panel d) on the spectral energy distribution. In all panels, the result of our fiducial model obtained for $l_{\rm nth}=30$, $l_B=10$, $\Gamma_{\rm inj}=3$, $\gamma_{\rm min}=1$, $\gamma_{\rm max}=1000$, $\tau_{\rm i}=2$, $R=5 \times 10^7$ cm is shown in red. Keeping all the other parameters constant, panel (a) shows the results for $l_{B}=1$ (green) and $l_B$=100 (blue). Panel (b) shows the spectra for  $\Gamma_{\rm inj}$=2.5 (green) and 3.5 (blue). Panel (c)  presents the results for $\tau_{\rm i}=5 \times 10^{-2}$ (pink), 0.5 (green) and 5 (blue). Finally panel (d) displays the spectra for $R=5 \times 10^6$ (green), $5\times 10^8$ (blue), and $5\times 10^{13}$ cm (cyan).
 }
 \label{fig:allpar}
\end{figure*}

\section{Simulation Results}\label{sec:simres}

In this section we present the incidence of the various parameters of our model (namely  $l_{\rm nth }$, $l_{\rm s}$, $l_{\rm c}$, $l_B$, $\tau_{\rm i}$, $R$, $\Gamma_{\rm inj}$, $\gamma_{\rm min}$ and $\gamma_{\rm max}$) on the shape of the electron and photon distribution. In Section~\ref{sec:nonthe} we explore models in which synchrotrons photons are the only source of soft seed photons ($l_{\rm s}=0$) and the protons are cold ($l_{\rm c}=0$) and the corona is powered only through non-thermal acceleration. { In Section~\ref{sec:coulombboiler} we illustrate and discuss the relative importance of $e$-$e$ Coulomb and Synchrotron self-absorption as thermalising mechanisms.} Then in Section~\ref{sec:ls}  
 we will investigate the effects of additional external soft-photons. Finally in Section~\ref{sec:effectsofCoulombheating} we illustrate the effects of Coulomb heating by hot protons. 
 The parameters and the results of the simulations discussed in this section are summarised in Table~\ref{tab:simus}.

\begin{table*}
 \centering
  \caption{Parameters of the model shown in Fig.~\ref{fig:le} to  \ref{fig:lcoul}, as indicated in the first column. The next 9 columns give the model parameters  as described in the text. The blackbody temperature $kT_{\rm bb}$ is given in eV, $R$ in cm.  $\alpha$ is the resulting spectral slope in the 2-10 keV range. $E_{\rm peak}$ is the maximum of the $\nu F_{\nu}$ spectral energy distribution (in keV). $\tau_{\rm T}$ is the total optical depth (i.e. including pairs). $kT_{\rm e}$ is the temperature of the thermal component of the electron distribution (in KeV) as determined from a fit with a Maxwellian. $kT_{\rm i}$ is the temperature of the protons in MeV.  $<\gamma>$  is the average Lorentz factor of the electron distribution.}
\begin{tabular}{c|cccccccc|ccccccccc}
\hline

 Fig. & $l_{\rm nth}$ & $l_{\rm c}$& $l_{\rm s}$& $l_{B}$& $\tau_{\rm i}$& $kT_{\rm bb}$& $R$& $\Gamma_{\rm inj}$& $\gamma_{\rm min}$&  $\alpha$& $E_{\rm peak}$& $\tau_{\rm T}$& $kT_{\rm e}$& $kT_{\rm i}$& $<\gamma>$ \\

\hline

 1,4 & 0.30 & 0 & 0 &  0.1 & 2 &   & 5 $\times$  $10^{7}$ & 3  & 1 &     0.76 &    65.70 &     2.00 &    44.97 &       &     1.15 \\
  1  &  3 & 0 & 0 & 1 & 2 &   & 5 $\times$  $10^{7}$ & 3 &  1 &  0.77 &    64.59 &     2.00 &    41.44 &     &     1.14  \\
1--5,8,9 & 30 & 0 & 0 & 10 & 2 &   & 5 $\times$  $10^{7}$ & 3 & 1&    0.75 &    65.08 &     2.02 &    35.24 &       &     1.14  \\
1 & 3 $\times$  $10^{2}$ & 0 & 0 &  $10^{2}$ & 2 &   & 5 $\times$  $10^{7}$ & 3 & 1&    0.70 &    55.71 &     2.79 &    20.57 &       &     1.10  \\
1,4 &3 $\times$  $10^{3}$ & 0 & 0 &  $10^{3}$ & 2 &   & 5 $\times$  $10^{7}$ & 3 &  1&   0.39 &    21.18 &     9.01 &     5.42 &        &     1.03   \\
\hline
2a & 30 & 0 & 0 & 1 & 2 &   & 5 $\times$  $10^{7}$ & 3 &   1 &  0.60 &    96.86 &     2.08 &    43.70 &     &     1.18   \\
2a & 30 & 0 & 0 &  $10^{2}$ & 2 &   & 5 $\times$  $10^{7}$ & 3 & 1 &    0.90 &    27.31 &     2.00 &    32.82 &    &     1.11 \\
2b & 30 & 0 & 0 & 10 & 2 &   & 5 $\times$  $10^{7}$ &     2.50 &   1 &  0.87 &    27.48 &     2.03 &    23.96 &   &     1.10 \\
2b & 30 & 0 & 0 & 10 & 2 &   & 5 $\times$  $10^{7}$ &     3.50 &    1&  0.65 &    92.35 &     2.01 &    45.54  &   &     1.17   \\
2c & 30 & 0 & 0 & 10 & 0.05 &   & 5 $\times$  $10^{7}$ & 3 &    1 &  0.98 & 0.017  &     0.25 & 118   &   &     1.80 \\
2c & 30 & 0 & 0 & 10 &     0.50 &   & 5 $\times$  $10^{7}$ & 3 &   1 &  0.83 &    92.58 &     0.64 &    93.56 &     &     1.46  \\
2c & 30 & 0 & 0 & 10 & 5 &   & 5 $\times$  $10^{7}$ & 3 &   1 &  0.64 &    28.09 &     5.00 &    12.33 &     &     1.04   \\
2d & 30 & 0 & 0 & 10 & 2 &   & 5 $\times$  $10^{8}$ & 3 &   1 &  0.76 &    64.19 &     2.02 &    34.37 &      &     1.13  \\
2d & 30 & 0 & 0 & 10 & 2 &   & 5 $\times$  $10^{6}$ & 3 &   1 &  0.74 &    66.24 &     2.02 &    36.19 &       &     1.14  \\
2d & 30 & 0 & 0 & 10 & 2 &   & 5 $\times$  $10^{13}$ & 3 &   1 &  0.78 &    61.45 &     2.02 &    30.01 &      &     1.12   \\
\hline
3 & 30 & 0 & 0 & 10 & 2 &   & 5 $\times$  $10^{7}$ & 3 & 1.5& 0.76 & 64.47 & 2.02& 34.29 & &1.13  \\% 0.75 &    65.08 &     2.02 &    35.24 &       &     1.14  \\
3 & 30 & 0 & 0 & 10 & 2 &   & 5 $\times$  $10^{7}$ & 3 & 2&    0.78  & 57.87 & 2.02 & 31.20&  & 1.13 \\% 0.75 &    65.08 &     2.02 &    35.24 &       &     1.14  \\
3 & 30 & 0 & 0 & 10 & 2 &   & 5 $\times$  $10^{7}$ & 3 & 3&    0.80   & 51.95 & 2.03 &  27.06 &  &1.12 \\ % 0.75 &    65.08 &     2.02 &    35.24 &       &     1.14  \\
3 & 30 & 0 & 0 & 10 & 2 &   & 5 $\times$  $10^{7}$ & 3 & 5&   0.82    &  37.58 & 2.04 & 20.38  &  &1.09 \\ % 0.75 &    65.08 &     2.02 &    35.24 &       &     1.14  \\
3 & 30 & 0 & 0 & 10 & 2 &   & 5 $\times$  $10^{7}$ & 3 & 10& 0.82 & 19.67     & 2.05 & 13.59   &  &1.07 \\ % 0.75 &    65.08 &     2.02 &    35.24 &       &     1.14  \\
\hline
 4$^\star$& 0.30 & 0 & 0 &  0.1 & 2 &   & 5 $\times$  $10^{7}$ & 3  & 1 & 0.77 & 59.45 & 2.00 & 23.41 & & 1.13 \\% &     0.76 &    65.70 &     2.00 &    44.97 &       &     1.15 \\
 4,5$^\star$& 30 & 0 & 0 & 10 & 2 &   & 5 $\times$  $10^{7}$ & 3 & 1  &   0.75  & 64.47  & 2.02 & 25.33& & 1.13 \\ %&    0.75 &    65.08 &     2.02 &    35.24 &       &     1.14  \\
 4$^\star$ & 3 $\times$  $10^{3}$ & 0 & 0 &  $10^{3}$ & 2 &   & 5 $\times$  $10^{7}$ & 3 &  1& 0.40 & 19.67 & 8.98 & 4.97 & & 1.03  \\%&   0.39 &    21.18 &     9.01 &     5.42 &        &     1.03   \\
 5$^\dagger$& 30 & 0 & 0 & 10 & 2 &   & 5 $\times$  $10^{7}$ & 3 & 1 & 1.01 & $3.8 \times 10^{-2}$ & 2.00 & 16.83 & & 1.07\\ 
 5$^\diamond$& 30 & 0 & 0 & 10 & 2 &   & 5 $\times$  $10^{7}$ & 3 & 1& 0.84 &  41.87 & 2.04 & 25.82 &  & 1.11\\ 
 5$^\bullet$& 30 & 0 & 0 & 10 & 2 &   & 5 $\times$  $10^{7}$ & 3 & 1& 0.83 & 41.86 &2.05 &   & & 1.10\\ 
\hline
6 & 5 & 0 & 15 & 10 & 2 & 420 & 5 $\times$  $10^{7}$ & 3 & 1 &    2.39 &     1.91 &     2.00 &     7.93 &     &     1.03 \\
6,7,8   & 15 &  0   & 15 & 10 & 2 & 420  & 5 $\times$  $10^{7}$ & 3 & 1 &    1.60 &     2.13 &     2.00 &    14.56 &     &     1.05   \\
6 & 50 & 0 & 15 & 10 & 2 & 420 & 5 $\times$  $10^{7}$ & 3 &   1 &  1.08 &     2.52 &     2.07 &    22.64 &       &     1.10  \\
6 & 5 & 0 & 15 & 10 &     0.2 & 420 & 5 $\times$  $10^{7}$ & 3 &   1 &  2.22 &     1.71 &     0.20 &    63.41 &       &     1.29  \\
6 & 15 & 0 & 15 & 10 &     0.2 & 420 & 5 $\times$  $10^{7}$ & 3 &  1 &   1.75 &     1.73 &     0.27 &    86.73 &      &     1.46 \\
6 & 50 & 0 & 15 & 10 &     0.2 & 420 & 5 $\times$  $10^{7}$ & 3 &   1 &  1.27 &     1.93 &     0.70 &    64.46 &       &     1.37 \\
\hline 
7 & 15 & 0 & 15 & 0 &        2 & 420 & 5 $\times$  $10^{7}$ & 3 &  1  &  1.57 &     2.14 &     2.02 &    14.87 &       &     1.06  \\
7 & 15 & 0 & 15 & 50 &      2 & 420 & 5 $\times$  $10^{7}$ & 3 &   1 &  1.59 &     2.14 &     2.00 &    15.29 &       &     1.05  \\
7 & 15 & 0 & 15 & 0 &     0.2 & 420 & 5 $\times$  $10^{7}$ & 3 &   1 &  1.76 &     1.79 &     0.41 &    59.26 &       &     1.37  \\
7 & 15 & 0 & 15 & 10 &     0.2 & 420 & 5 $\times$  $10^{7}$ & 3 &  1 &   1.75 &     1.73 &     0.27 &    86.73 &      &     1.46  \\
7 & 15 & 0 & 15 & 50 &     0.2 & 420 & 5 $\times$  $10^{7}$ & 3 &   1 &  1.71 &     1.71 &     0.21 & 117 &  &     1.51\\
\hline
8 &     0.30 & 0 &    29.7 & 10 & 2 & 499 & 5 $\times$  $10^{7}$ & 3 & 1 &    4.50 &     2.00 &     2.00 &     2.02 &    &     1.01  \\
8 & 3 & 0 & 27 & 10 & 2 & 487 & 5 $\times$  $10^{7}$ & 3 &   1 &  3.28 &     2.04 &     2.00 &     3.48 &       &     1.01   \\
8 & 9 & 0 & 21 & 10 & 2 & 457  & 5 $\times$  $10^{7}$ & 3 &   1 &  2.17 &     2.11 &     2.00 &     8.48 &      &     1.03   \\
8 & 21 & 0 & 9 & 10 & 2 & 370 & 5 $\times$  $10^{7}$ & 3 &   1 &   1.23 &      2.10 &     2.01 &    21.82 &       &     1.08   \\
8 & 27 & 0 & 3 & 10 & 2 & 281  & 5 $\times$  $10^{7}$ & 3 &   1 &   0.91 &    47.05 &     2.01 &    30.34 &     &     1.12  \\
\hline 
9 & 0 & 30 & 0 & 10 & 2 &   & 5 $\times$  $10^{7}$ &   &    & 0.26 & 288  &     2.15 & 134 &    9.34 &     1.47   \\
9 & 1 & 29 & 0 & 10 & 2 &   & 5 $\times$  $10^{7}$ & 3 &  1 &   0.44 & 182 &     2.00 &    87.65 &      5.37 &     1.29  \\
9 & 10 & 20 & 0 & 10 & 2 &   & 5 $\times$  $10^{7}$ & 3 & 1 &    0.64 &    93.82 &     2.01 &    53.76 &     1.80 &     1.18  \\
9 & 15 & 15 & 0 & 10 & 2 &   & 5 $\times$  $10^{7}$ & 3 &  1 &   0.68 &    89.71 &     2.01 &    47.72 &    1.13 &     1.16   \\
9 & 20 & 10 & 0 & 10 & 2 &   & 5 $\times$  $10^{7}$ & 3 &  1 &   0.71 &    69.73 &     2.01 &    42.98 &      0.65 &     1.15 \\
\hline
\end{tabular}
\flushleft $^\star$ $e$-$e$ Coulomb neglected\\
$^\dagger$ synchrotron heating neglected\\
$^\diamond$ synchrotron heating and cooling neglected\\
$^\bullet$ $e$-$e$ Coulomb , synchrotron heating and cooling neglected\\
\label{tab:simus}
\end{table*}

\subsection{Synchrotron self-Compton models with pure non-thermal injection}\label{sec:nonthe}

In this section we assume that the protons are cold ($l_{\rm c}=0$) and the external photons are neglected ($l_{\rm s}=0$).
Figure~\ref{fig:le} shows the dependence of the photon spectrum on $l_{\rm nth}$ for  $l_B=l_{\rm nth}/3$, which corresponds to approximate equipartition  of magnetic field with radiation.  
The other fixed parameters are $\Gamma_{\rm inj}=3$, $R=5 \times 10^7$ cm, and $\tau_{\rm i}=2$.
For a wide range of compactness the spectrum peaks around 65 keV and and the X-photon index is $\Gamma\simeq1.7$ (see also Table~\ref{tab:simus}).  { The right panel of Fig~\ref{fig:le} shows the steady state particle distribution corresponding to the these photon spectra. These particle distributions are all  thermal at low energies with a non-thermal tail at high energies.  
The non-thermal electrons produce self-absorbed synchrotron radiation peaking at the synchrotron turn-over frequency  at photon energies of a few eV.  The hybrid thermal non-thermal distribution then Compton up-scatters the self-absorbed synchrotron radiation, forming spectra extending into the hard X-ray and $\gamma$-ray domain.  These spectra are dominated by the Comptonisation by the thermalised particles. The peak and energy of the cut-off depends on their temperature.} In the following we will refer to this process as Synchrotron Self-Compton emission (or SSC). 

In terms of particle kinetics, those results can be understood as follows. The injected high energy electrons cool down by emitting { SSC radiation and by interacting through Coulomb collisions with lower energy particles. 
We define  $\gamma_t$ as the Lorentz factor corresponding to the energy of the electrons radiating most of their synchrotron luminosity around the synchrotron turn over frequency. Below the critical Lorentz factor $\gamma_{\rm t}$, the particles emit synchrotron in the self-absorbed regime: their emission is immediately absorbed by lower energy electrons. This very fast process of synchrotron emission and self absorption represents a very efficient and fast thermalising mechanism. Very quickly the low energy part of the electron distribution forms a quasi Maxwellian. This synchrotron boiler effect is able to transfer  a large fraction of the non-thermal energy of the electrons (of Lorentz factor $<\gamma_t$) into heating of the lowest energy particles, keeping the effective temperature around 40 keV. 

 An analytical estimate of $\gamma_t$ is derived in Appendix~\ref{sec:appendix}.  This gives $\gamma_t\simeq10$, the exact value being rather insensitive to the model parameters, see equations~(\ref{eq:gtrad}) and (\ref{eq:gtcoul}). Not only the electrons at or above  $\gamma_{\rm t}$ do not contribute to the heating of the lower energy particles but their soft synchrotron radiation also provides the dominant contribution to the Compton cooling of the plasma. Therefore the efficiency of the synchrotron boiler as a heating mechanism will depend strongly on the relative fraction of electrons present  above and below $\gamma_{\rm t}$.}
\begin{figure*}
 \includegraphics[width=\textwidth]{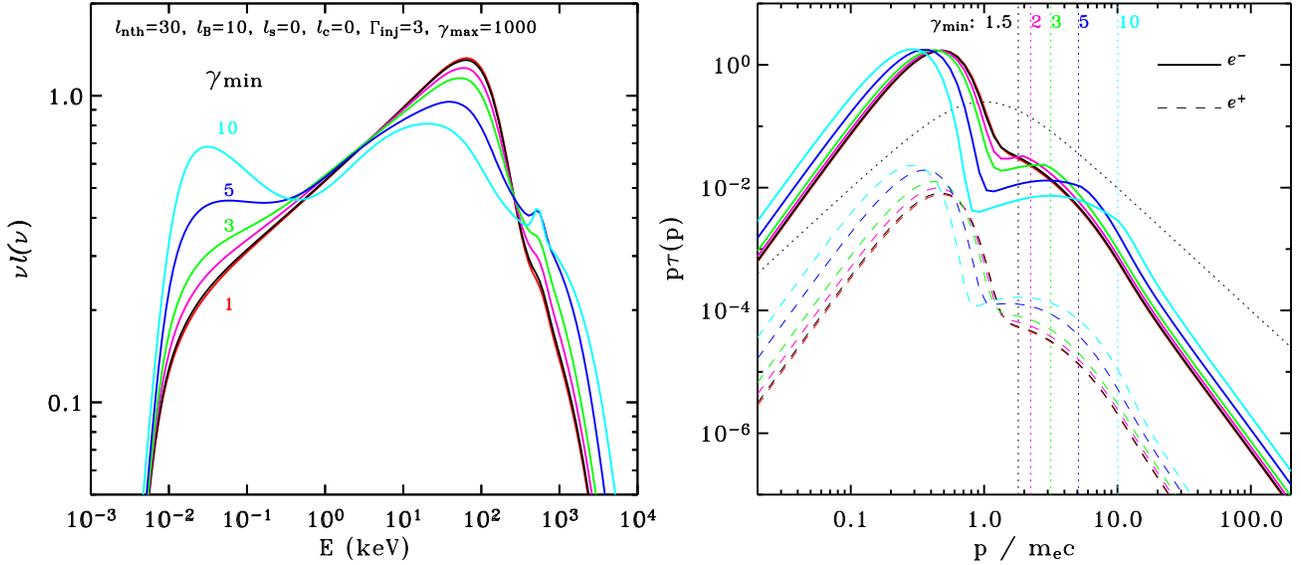}
 \caption{Effects of varying $\gamma_{\rm min}$ in pure synchrotron self-Compton models. Simulated photon spectra (left panel) and particle distributions for $\gamma_{\rm min}$= 1 (red), 1.5 (black), 2 (pink),  3 (green),  5 (blue) and 10 (cyan).  The other parameters are $l_{\rm nth}=30$, $l_{B}=10$, $\Gamma_{\rm inj}=3$, $\gamma_{\rm max}=1000$, $\tau_{\rm i}=2$, $R=5\times 10^7$ cm. In the right hand side panel the full curves show the electron while the dashed ones show the positron distributions. { The dotted curve shows the shape of the injected electron distribution.}}
 \label{fig:gmin}
\end{figure*}
{ At the same time Coulomb collisions between high energy and low energy electrons also plays an important (and possibly dominant)  thermalising role, similar to that of the synchrotron boiler. The relative importance of synchrotron and Coulomb boilers will be discussed in Section~\ref{sec:coulombboiler}}.

{ At fixed $l_{B}/l_{\rm nth}$ both synchrotron and Compton losses scale like $l_{\rm nth}$ (see equation~\ref{eq:lsync}) and the injected power also scales like $l_{\rm nth}$.  The spectra then depend only weakly on compactness. At low compactness the small differences between the  spectra of various compactness are entirely due to Coulomb effects and  will be discussed in Section~\ref{sec:coulombboiler}}.
At very large $l_{\rm nth}$, pair production effects become important and electron positron pairs contribute significantly to the total optical depth ($\tau_{\rm T}\simeq9$ for $l_{\rm nth}=3 \times 10^3$). As a consequence the effective temperature decreases, as more particles share the boiler energy, but the Compton parameter actually increases producing even harder spectra ($\Gamma$=1.39 for  $l_{\rm nth}=3 \times 10^3$). We also note that, at high compactness, pair annihilation leads to the formation of the line apparent around 511 keV  in some spectra of Fig.~\ref{fig:le}.

These simulations show that  even though the electron injection is purely non-thermal, most of the radiated luminosity emerges as quasi-thermal Comptonisation. The basic spectral properties of black hole binaries in the LHS appear to be produced over a wide range of luminosity (at least up to 0.3$ L_{\rm E}$) through pure non-thermal electron injection in a magnetised plasma, with no need for additional heating mechanism nor soft photons.

\begin{figure*}
 \includegraphics[width=\textwidth]{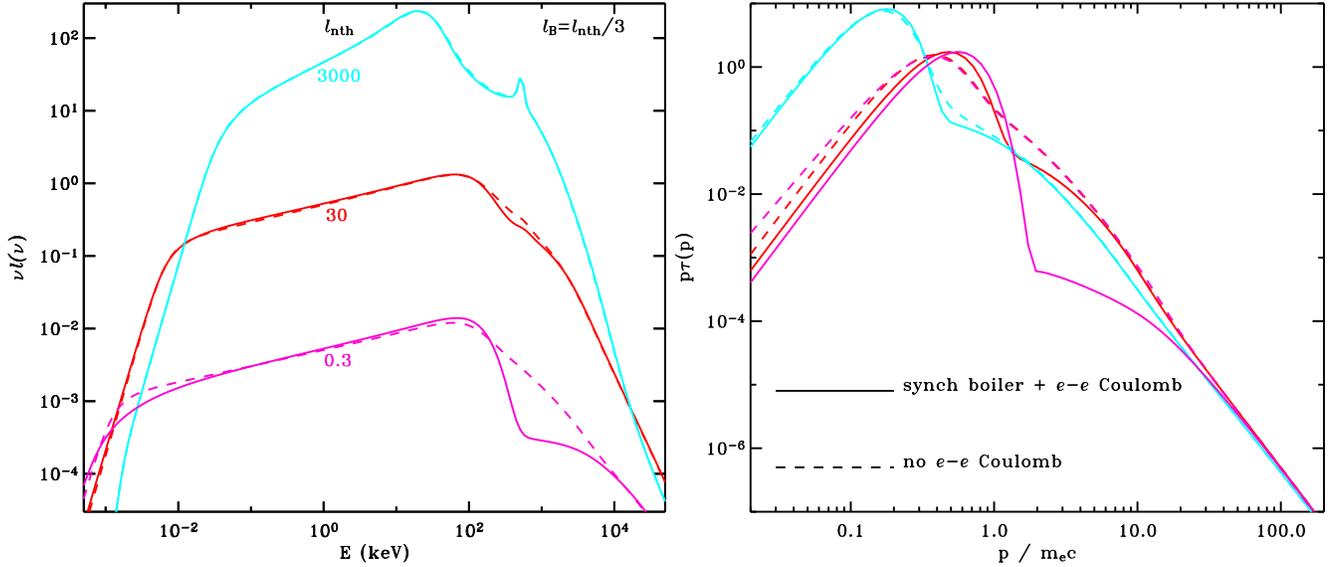}
 \caption{{ Synchrotron self-Compton models with pure non-thermal injection: effects of neglecting $e$-$e$ Coulomb.     
Simulated photon spectra (left panel) and lepton distributions (right panel) for $l_{\rm nth}$= 0.3 (pink), 30 (red),  3000 (cyan) and $l_{\rm nth}/l_{B}=3$ (same parameters as in Fig~\ref{fig:le}). The full curves show the results of the full calculation. The dashed curves show the results when $e$-$e$ Coulomb collisions are switched-off.}}
 \label{fig:lewocoul}
\end{figure*}

Let us now focus on the  fiducial case $l_{\rm nth}=30$ ($L\simeq 0.03 L_{\rm E}$) and evaluate the effects of the other parameters. As shown in panel $(a)$ of Fig~\ref{fig:allpar}, increasing $l_{B}$ increases the soft synchrotron emission. As a consequence, the competing process, i.e. inverse Compton emission of the non-thermal electrons, is reduced.  Simultaneously, the temperature of the thermal component is reduced due to the enhanced synchrotron soft cooling flux. This leads to softer X-ray spectra, peaking at lower energy. 
In the limit of vanishing $l_{B}$ the electrons cannot radiate efficiently their energy through SSC. Other processes such as bremsstrahlung (which is otherwise negligible in the situations considered in this paper) and its Comptonisation  may become the dominant cooling mechanism of the plasma. The $l_B=0$ temperature represents  the maximum temperature achievable with pure non-thermal injection. 

Increasing $\Gamma_{\rm inj}$ decreases { the fraction of electrons with Lorentz factor larger than  $\gamma_{\rm t}$, therefore the self-absorbed synchrotron emission is reduced  and the energy available for the synchrotron boiler is increased.  The temperature of the Maxwellian component is then higher. }
This leads  to harder Comptonised spectra peaking at higher energy, with a weaker non-thermal fraction, as shown on panel $(b)$ of Fig.~\ref{fig:allpar}. 

Increasing $\tau_{\rm i}$ decreases the average energy of the thermal component (since more particles have to share the same amount of energy), this reduces the peak energy of the Comptonised spectrum (see  panel $(c)$ of Fig.~\ref{fig:allpar}). 
Simultaneously, {  at larger $\tau_{\rm i}$ the radiation energy density is larger and, as a consequence, the Compton losses of the non-thermal leptons are also larger at the expense of the synchrotron emission.  The synchrotron self-absorbed emission is therefore weaker (see equation~(\ref{eq:lsync})) and, in order  to radiate the same luminosity, the Comptonised X-ray spectrum must be harder.}

As shown on panel $(d)$ of Fig.~\ref{fig:allpar}, varying the size of the emitting region $R$ has very little effects on the photon spectrum, at least at high energy. The main effect is to shift the cyclotron and the self absorption frequencies. { The self absorption frequency $\epsilon_{\rm t}$ scales like  $R^{-b}$ with $b=\frac{3+\Gamma_{\rm inj}}{10+2\Gamma_{\rm inj}}$ see equations~(\ref{eq:ec}) and (\ref{eq:et1})}. The soft cooling synchrotron photon compactness  however remains nearly constant { ($l_{\rm sync}\propto R^{c}$ with $c=\frac{2-\Gamma_{\rm inj}}{10+2\Gamma_{\rm inj}}$, see equation~(\ref{eq:lsync}))}, which ensures the temperature of the thermal component does not change significantly. When $R$ increases, the spectrum has to be slightly softer for energy conservation reasons. Overall the effects are almost negligible except for huge variations of $R$. Panel $(d)$ of Fig.~\ref{fig:allpar} shows a spectrum computed for a size relevant to AGN ($R=5\times 10^{13}$ cm). The spectrum is softer and peaks at slightly lower energy but overall the shape is very similar. This kind of model would produce LHS like spectra in AGN. 

{ Fig.~\ref{fig:gmin} shows the effects of varying the minimum Lorentz factor of the injected electrons. As long as $\gamma_{min} < 2$ the exact value has virtually no effects on the steady state particle and photon distribution. For $\gamma_{\rm min}>2$, on the contrary,  the synchrotron emission increases quickly with $\gamma_{\rm min}$, cooling down the plasma and producing softer hard X-ray spectra.  This effect is simply due to the fact that we keep $l_{nth}$ constant when increasing $\gamma_{\rm min}$. Since less power is injected at low energies, this involves increasing the normalisation of the injected distribution and therefore increasing the non-thermal particle density at energies above $\gamma_{\rm t}$, and therefore the synchrotron luminosity. This increase in normalisation is not linear with $\gamma_{\rm min}$. For $\gamma_{min}=2$ the normalisation of the injected distribution is only about 30 percent larger than for $\gamma_{\rm min}=1$, and the effect is weak. For larger $\gamma_{\rm min}$ the normalisation has to increase faster. For $\gamma_{\rm min}$=4 and 10  it is larger by factors of about  2 and 5 respectively. The increased synchrotron emission has then a strong impact on the energy balance in the corona.}

\begin{figure*}
 \includegraphics[width=\textwidth]{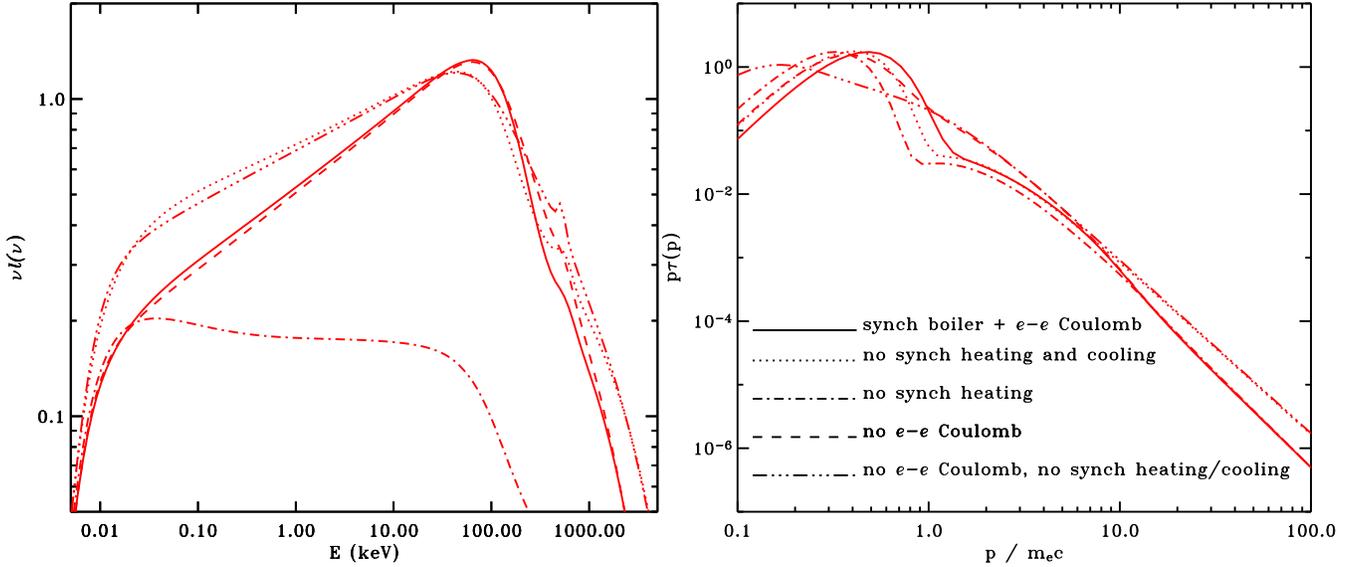}
 \caption{Comparison of the effects of the synchrotron and $e$-$e$ Coulomb boilers in a SSC model with pure non-thermal injection. Simulated photon spectra (left panel) and electron distributions (right panel) for the full calculation (full curve), neglecting the effects  of synchrotron heating on the particle distribution (dot-dash), neglecting both synchrotron heating and cooling (dotted curve), neglecting $e$-$e$ Coulomb (dashed curve) and neglecting simultaneously Coulomb, synchrotron heating and cooling (3 dots-dash).  The parameters are those of our fiducial model (i.e. $l_{\rm nth}=30$, $l_{B}=10$, $\Gamma_{\rm inj}=3$, $\tau_{\rm i}=2$, $R=5 \times 10^7$ cm).}
 \label{fig:lh30}
\end{figure*}

\subsection{Synchrotron boiler vs $e$-$e$ Coulomb boiler}\label{sec:coulombboiler}

{ Since the effects of synchrotron absorption on the electron distribution have been routinely neglected in previous works, it is interesting to check their importance and compare them with other thermalising processes.
Another important thermalising mechanism is the $e$-$e$ Coulomb interaction which may keep the particle distribution close to a Maxwellian at low energies and  transfer a fraction of the power from high to lower particle energies. 
  
 The collisions between electrons/positrons and ions also contribute to the thermalisation process, however the effects are important only when the temperature of the ions is larger than that of the electrons (this case will be considered in Section~\ref{sec:effectsofCoulombheating}). If the temperature of ions and electrons is comparable we found the effects completely negligible compared to $e$-$e$ collisions (and the $e$-$p$ collisions were actually neglected in the simulations with $l_{\rm c}=0$ presented in this paper).

In principle bremsstrahlung emission and its self-absorption could also play a role similar  to that of the synchrotron boiler. However,  at least in the case of $e$-$p$ bremsstrahlung, we found that the thermalising effects are fully negligible in the parameter regime considered in this work and we decided not take $e$-$p$ bremsstrahlung  into account in the simulations presented in this paper. $e$-$e$~bremsstrahlung is not taken into account in our code but its effects are also expected to be weak (see discussion in Belmont et al., 2008b).

Therefore the only important thermalising mechanism besides synchrotron self-absorption is $e$-$e$ Coulomb. In terms of energy loss by non-thermal electrons the ratio of Coulomb losses to synchrotron losses $\dot\gamma_{\rm  c}/\dot\gamma_{s}\propto (\tau_{\rm T}/l_{B})\beta^{-3}\gamma^{-2}$ (see Appendix~\ref{sec:appendix}).  Therefore Coulomb collisions dominate over synchrotron losses at low magnetic compactness and low particle energies. And for this reason unless the ratio $l_{\rm B}/\tau_{\rm T}$ is very large, the Coulomb process is  the  dominant process  keeping the low energy part of the lepton distribution thermal. The synchrotron boiler is the dominant thermaliser only for $l_{B}/\tau_{\rm T}\ga100$ (see discussion in GHS98). However since synchrotron losses dominate at higher particle energies, the  synchrotron boiler can be more effective at transferring the energy from the non-thermal  to the thermal particles (i.e. as a heating mechanism).
In practice, the synchrotron boiler  contributes significantly to the heating of the Maxwellian distribution for  $l_{B}/\tau_{\rm T}\ga0.2$  (see equation~(\ref{eq:syncboilrules})).

Using the results from our code, we can estimate the synchrotron and Coulomb heating power.   We define it as the net  power transferred from particles of Lorentz factor above $\gamma=1+10kT_{\rm e}/m_{\rm e}c^2$ to particles of lower energy.  Considering the models of Fig.~\ref{fig:le}, we find that for $l_{B}=0.1$ the synchrotron heating is negligible compared to Coulomb (in terms of compactness the Coulomb and synchrotron heating are 0.19, and 3$\times 10^{-3}$ respectively). The model with $l_{B}=1$ is just above the threshold for which synchrotron heating becomes significant. For this model, we find 1.48 and 0.14 respectively for the  Coulomb and synchrotron heating compactness.  In the model with $l_{B}=10$ they are comparable (5.9 and 4 for Coulomb and synchrotron heating respectively). At $l_{B}$=100, heating is dominated by synchrotron (we find 26 and 59 for the Coulomb and synchrotron heating compactness respectively). Finally for $l_{B}=1000$, Coulomb effects are very weak (we find 44 and  314  for the Coulomb and synchrotron heating compactness respectively).

\begin{figure*}
 \includegraphics[width=\textwidth]{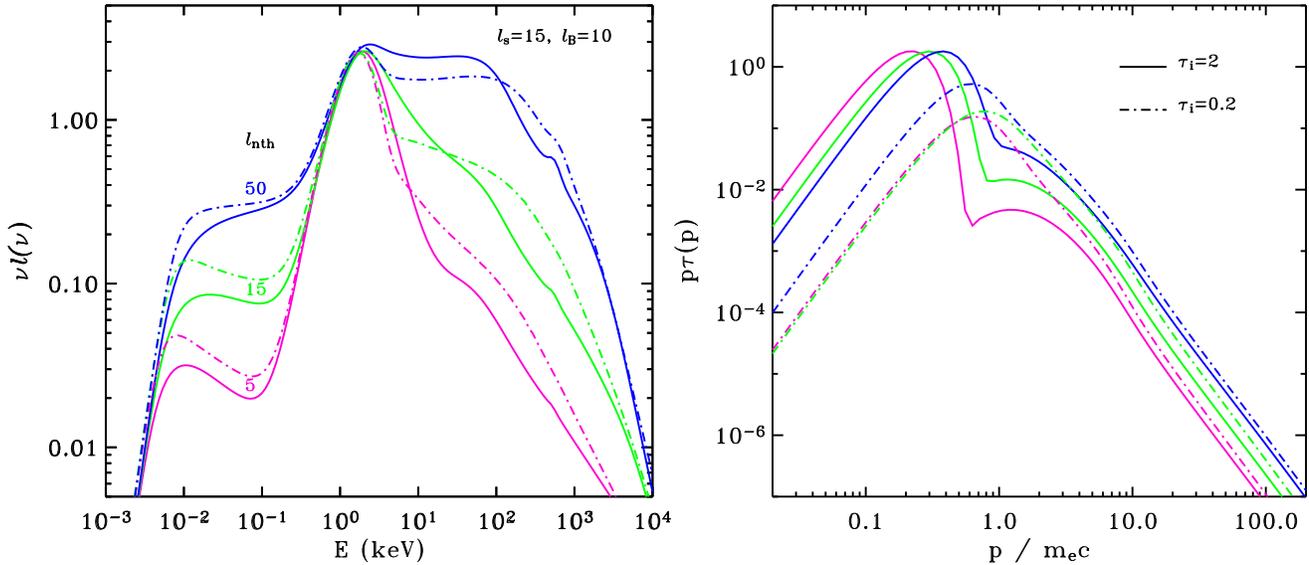}
 \caption{{ Non-thermal injection models dominated by external photons.} Effects of varying the non-thermal compactness $l_{\rm nth}$ at constant $l_{\rm s}=15$  and $l_B=10$.
    on the photon spectra  (left hand side) and the lepton distributions (right hand side).  The pink, green and blue curves stand for $l_{\rm nth}$=5, 15 and 50 respectively. The full curves are models with Thomson depth $\tau_{\rm i}$=2, while the dot-dashed curves show the results for $\tau_{\rm i}$=0.2. 
 The other parameters are  $kT_{\rm bb}=420$ eV, $\Gamma_{\rm inj}=3$, $\gamma_{\rm min}=1$, $\gamma_{\rm max}=1000$, $R=5 \times 10^7$ cm, see Table~\ref{tab:simus}.}
 \label{fig:lnthlsoft}
\end{figure*}

Besides thermal heating,  Coulomb interactions also have a significant effect on the non-thermal tail of the lepton energy distribution.  
Indeed, for the non-thermal particles of lowest energy,
Coulomb cooling usually  dominates over Compton and synchrotron losses.
 This tends to deplete faster the electrons from this energy range. As a result  the electron distribution is less steep than the expected $\Gamma_{\rm inj}$+1 slope obtained in the case of pure radiative losses (see equation~(\ref{eq:analdistrib})). This flattening occurring in the range $p=$1--10, can be observed in the electron distributions presented in Fig.~\ref{fig:le}. This effect is also illustrated by Fig.~\ref{fig:lewocoul} which compares particles and photons spectra of the simulations with  $l_B$=0.1, 10 and 1000 of Fig.~\ref{fig:le}  with models computed with the $e$-$e$ Coulomb collisions switched-off.  
 
Because the importance of Coulomb losses decreases with compactness, the deviation from a power law starts at higher electron energies for lower compactness.  An estimate for the  momentum corresponding  to the transition from radiative to Coulomb cooling  is provided by equation~(\ref{eq:pc}). For the simulations with $l_B$=0.1, 10 and 1000 this gives   $p_{\rm c}$=19.4, 1.98 and 0.56  respectively.

The main effects of Coulomb collisions on the photon spectrum  is to increase the sharpness of the thermal comptonisation cut-off by depleting the supra-thermal particles. This effect is very significant, and should be taken into account when comparing to data. For instance,  in section~\ref{sec:lhscyg},  we compare SSC models with a LHS spectrum of Cygnus X-1 and find a good agreement when Coulomb cooling is included while no reasonable fit could be found neglecting this effect. As seen  on Fig.~\ref{fig:lewocoul}, the effects of Coulomb collisions on the spectral shape decrease at higher compactness as discussed above.    For the  high compactness case $l_B=1000$, the effects of Coulomb collisions are practically negligible, both thermalisation and thermal heating are dominated by the synchrotron boiler.  

In addition, as expected, the temperature of the thermal electrons increases when Coulomb is taken into account (see Table~\ref{tab:simus}).  As a consequence, the $\nu F_\nu$ spectra peak at slightly higher energies in the simulations including  Coulomb. The decrease of the temperature with compactness in the simulations of Fig~\ref{fig:le} is related to the dependence of Coulomb heating on $l_B$. 
In the simulations without Coulomb, the lower temperature of the Maxwellian is compensated by the much larger density of supra-thermal particles which also contributes to the Comptonisation process. As the average energy of the Comptonising particles is almost identical,  the X-ray spectral slopes are not strongly affected by the Coulomb collisions. 

Moreover, we note that the Coulomb and synchrotron boilers are competing to produce the same result, when Coulomb collisions are neglected the synchrotron boiler takes over transferring a large fraction of the non-thermal energy that would have otherwise been transferred via Coulomb collisions to the thermal pool of electrons. In the case, $l_{B}=0.1$, the synchrotron heating power is $7.2 \times 10^{-2}$, i.e. larger by a factor 24, when the Coulomb boiler is neglected. In our fiducial model $l_{B}=10$ the synchrotron heating power doubles to become 7.4, compensating almost fully for the lack of Coulomb heating (the total heating compactness was about 10 in the full calculation).

In order to understand better the respective roles of Coulomb and synchrotron in the thermalisation and plasma heating processes, it is useful to see what happens when the synchrotron boiler is turned off.
Fig~\ref{fig:lh30} shows our fiducial simulation for $l_{\rm nth}$=30 in which the heating of the leptons due to synchrotron self-absorption is neglected (but heating by $e$-$e$ collisions is taken into account).
{ In this simulation, the energy of the absorbed  synchrotron photons is artificially lost and a large fraction of the injected energy is not radiated. As a consequence, the luminosity is lower by a factor of 4 with respect to the full calculation.}
 The temperature of the thermal leptons is then significantly lower (only 16 keV see Table~\ref{tab:simus}). As a consequence the spectra is much steeper, with photon index $\Gamma\simeq 2$,  incompatible with the typical slopes observed in the LHS. The heating of the thermal population by $e-e$ Coulomb collisions appears too weak to sustain the electron temperature observed in the hard state. 

However, when comparing to Coulomb process we should consider the net effect of the magnetic field on the lepton distribution and switch-off  both synchrotron heating and cooling. 
Fig~\ref{fig:lh30} shows the same simulation with not only the synchrotron heating but also synchrotron  cooling  turned-off. 
Then the temperature is higher ($kT_{\rm e}$=26 keV) than in the simulation including synchrotron cooling yet significantly lower than in the full calculation. However this is essentially  because the synchrotron losses of the high energy electrons are not taken into account and therefore the density of electrons above $\gamma_{\rm t}$ is larger by about a factor of 2. As a consequence, the self-absorbed synchrotron luminosity is also larger by a factor of two.  In this simulation the thermal Coulomb heating  compactness  is 8.6 i.e. comparable to total (Coulomb + synchrotron) heating rate of the full calculation ($\simeq10$). What makes the temperature lower is the larger  Compton  cooling rate of the thermal electrons. This  leads to a softer hard X-ray  photon index of $\Gamma\simeq1.8$. 
In fact the temperature of the thermal electrons is close to that obtained when neglecting $e$-$e$ Coulomb, however, because of the enhanced Compton cooling rate the average  Lorentz factor is lower  ($\langle\gamma\rangle$=1.11 when synchrotron heating and cooling are neglected, $\langle\gamma\rangle$=1.13 when Coulomb is neglected). 

Finally, we note that the results without Coulomb are close to those of  the full calculation. On the contrary, the photon spectra obtained  
 neglecting both synchrotron cooling and heating are similar to those obtained neglecting all thermalising processes (i.e. switching-off both Coulomb and synchrotron boilers also shown in Fig.~\ref{fig:lh30}). }

\subsection{Effects of external soft photons}\label{sec:ls}

\begin{figure*}
 \includegraphics[width=\textwidth]{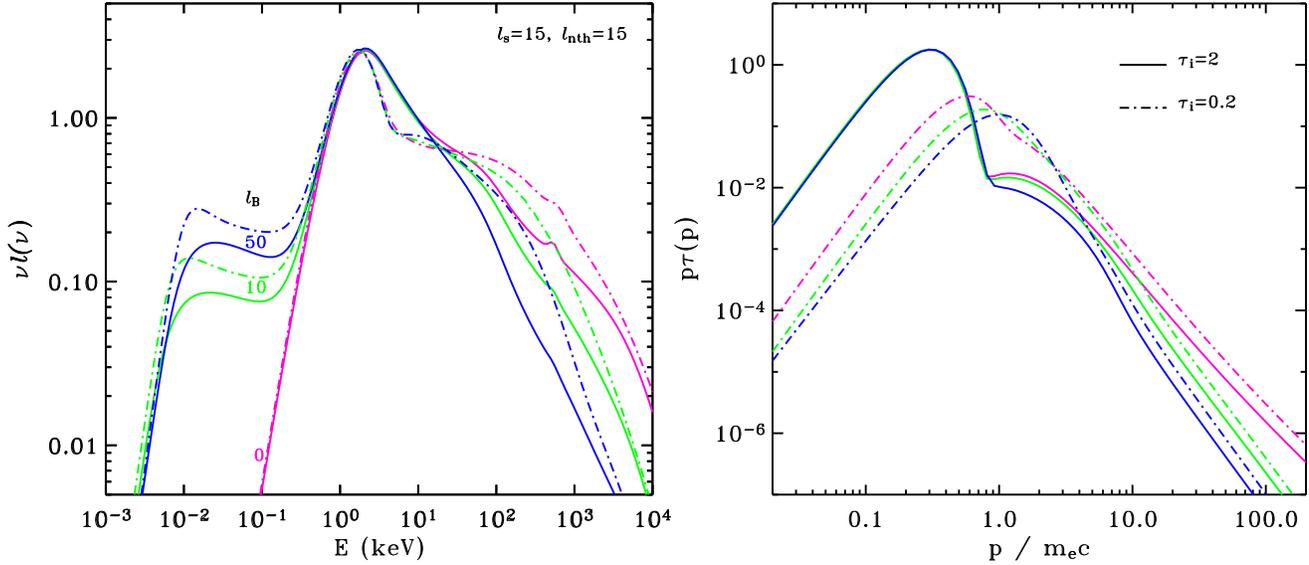}
 \caption{{ Non-thermal injection models dominated by external soft photons.} Effects of varying the magnetic compactness $l_{B}$ at constant $l_{\rm s}$=15 and $l_{\rm nth}$=15,
    on the photon spectra  (left hand side) and on the lepton distributions (right hand side).  The pink, green and blue curves stand for $l_{B}$=0, 10 and 50 respectively. The full curves are models with Thomson depth $\tau_{\rm i}$=2, while the dot-dashed curves show the results for $\tau_{\rm i}$=0.2. 
 The other parameters are  $kT_{\rm bb}=420$ eV, $\Gamma_{\rm inj}=3$, $\gamma_{\rm min}=1$, $\gamma_{\rm max}=1000$, $R=5 \times 10^7$ cm, see Table~\ref{tab:simus}.
 }
 \label{fig:lblsoft}
\end{figure*}

In addition to soft photons produced internally through synchrotron, the hot comptonising plasma may intercept a fraction of  the thermal radiation from the accretion disc.  We denote as $l_{\rm s}$ the compactness associated to this additional soft photon injection.

The main effect of the external soft photons field, is to increase the Compton cooling rate  of the leptons. The equilibrium temperature of the thermal component is lower than in pure SSC models. If $l_{\rm s}$ is strong ($l_{\rm s}\sim l_{\rm nth}$) the thermalised leptons are so cool that most of the luminosity is radiated by the non-thermal particles. Basically the effect of $l_{\rm s}$, is to increase the fraction of radiation emitted through non-thermal Comptonisation.    

At  fixed $l_{\rm s}$ and $l_B$ increasing $l_{\rm nth}$  increases the amount of energy available for the synchrotron boiler, the temperature of the plasma increases and the high energy emission is not only stronger, but also harder as can be seen  in Fig.~\ref{fig:lnthlsoft}. As shown on this figure and in Table~\ref{tab:simus}, the temperature is higher at lower optical depth because less particles share the same energy. This behaviour is similar to what obtained in the SSC model. In the SSC model this resulted in a softer spectrum at low optical depth (see Fig.~\ref{fig:allpar}c) while Fig.~\ref{fig:lnthlsoft} shows harder X-ray spectra at lower optical depth. This is because, as discussed in section~\ref{sec:nonthe}, in the SSC models the flux of seed photons for comptonisation is larger at small $\tau_{\rm T}$, whereas it is independent of $\tau_{\rm T}$ when $l_{\rm s}$ is important. 

We note that in presence of soft photons the effects of the  magnetic field are somewhat different from those of pure synchrotron self-Compton models. As discussed in Section~\ref{sec:nonthe} and illustrated in Fig.~\ref{fig:allpar}a, in absence of soft photons, increasing $l_{B}$ increases the radiative losses (both trough synchrotron and and synchrotron self-Compton), and causes the temperature of the plasma to decreases. When the losses are dominated by Compton on external soft-photons, then increasing $l_B$ has negligible effects on the radiative losses of  the plasma, yet this transfers a larger fraction of non-thermal energy to the lower energy particles which results in a higher plasma temperature.    
Therefore, when synchrotron self-Compton dominates, the magnetic fields tends to cool down
 the thermal component, when external Compton dominates it heats up this distribution.
 This effect however strongly depends on the Thomson optical depth of the plasma. For instance if we consider the simulation shown in Fig.~\ref{fig:lblsoft} with $l_B=10$, $l_{\rm s}$=15, and $l_{\rm nth}$=15 the equilibrium temperature of the thermal component is about 14.6 keV. Setting $l_{B}=0$, we obtain the same temperature, while setting $l_B=50$ we obtain $kT_{\rm e}=15.3$ keV, i.e. the temperature increases by only 5 percent. Then, repeating the same experiment with an optical depth reduced to $0.2$, we find a temperature of $kT_{\rm e}$=86 keV  for $l_B=10$  and  the temperature increases by a factor of 2 between the case $l_{B}=0$ ($kT_{\rm e}$=59 keV) and $l_{B}$=50 ($kT_{\rm e}$=117 keV). This is because, while the Compton and synchrotron losses are nearly independent of the optical depth,  the energy of the synchrotron boiler is shared among a smaller number of particles in the optically thin case, making the synchrotron heating much more effective. We also note that even when the optical depth is too thick for the  magnetic field to play a significant role in heating the thermal component of the plasma,  it significantly affects the emission of the non-thermal particles. For a stronger $l_B$ the high energy particles radiate their energy in the UV band rather than at  gamma-ray energies (see Fig.~\ref{fig:lblsoft}). Therefore increasing $l_B$ also reduces the fraction of high energy radiation emitted through non-thermal Comptonisation.

 \begin{figure*}
 \includegraphics[width=\textwidth]{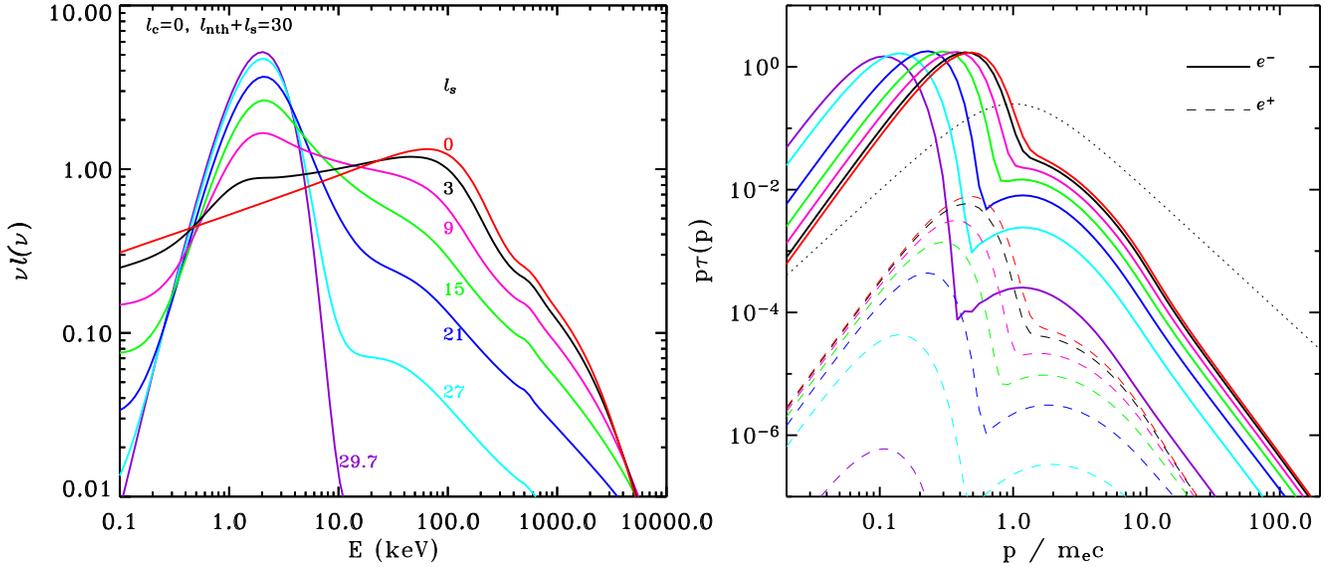}
 \caption{{ Pure non-thermal models with external soft photons.} Effects of varying the soft photon compactness $l_{\rm s}$ at constant $l$ on the photon spectrum  (left hand side) and the particle distributions (right hand side).  The red curve shows our fiducial model obtained for $l_{\rm s}=0$, $l_{\rm nth}=30$, $l_B=10$, $\Gamma_{\rm inj}=3$, $\gamma_{\rm min}=1$, $\gamma_{\rm max}=1000$, $\tau_{\rm i}=2$, $R=5 \times 10^7$ cm. The other curves show a sequence of models obtained  keeping $l=l_{\rm s}+l_{\rm nth}=30$ while the soft photon compactness $l_{\rm s}$ is varied. The black, pink, green, blue, cyan and purple curves show the results for $l_{\rm s}$=3, 9, 15 21, 27, 29.7 respectively.  The soft photons are injected with a blackbody spectrum of temperature $kT_{\rm bb}$=213.5 $l_{\rm s}^{1/4}$ eV. In the right hand side panel, the full curves show the electron while the dashed ones show the positron distributions.}
 \label{fig:lsoft}
\end{figure*}

Fig~\ref{fig:lsoft} shows a sequence of spectra, starting from our fiducial case $l_{\rm nth}=30$, $l_{B}=10$ in which  $l_{\rm s}$  is increased keeping the total compactness $l=l_{\rm s}+l_{\rm nth}$ constant. As the soft photons flux rises in the corona, inverse Compton increasingly dominates the cooling of the non-thermal particles eventually turning the synchrotron boiler off. As $l_{\rm s}$ is increased,  the electron distribution cools down to the Compton temperature. The emissivity of the thermal component is strongly reduced and the hard X-ray spectrum becomes gradually dominated by the emission of the cooling non-thermal distribution\footnote{ The secondary bumps observed in the electron distribution of some of the simulations presented in Fig.\ref{fig:lnthlsoft},~\ref{fig:lblsoft},~\ref{fig:lsoft} are essentially due to the use of the $p\tau(p)$ representation in which a power law distributions in $\gamma$ space has a maximum around $p=1$.}. Then, the non-thermal emission vanishes as $l_{\rm nth}$ decreases. We note that this result as well as the main results of Section~\ref{sec:nonthe} are qualitatively similar to those obtained independently by Vurm and Poutanen (2008).

This spectral sequence, where $l_{\rm s}$ is increased,
is reminiscent of that observed during the state transition of accreting black holes (DGK07; Del Santo et al. 2008).  This indicates that cooling by the soft photons from the accretion disc probably plays a major role in the changing appearance of the corona during such events.  It was known from spectral fits with {\sc eqpair} that due to the strong thermal emission from the disc, the ratio $l_{\rm h}/l_{\rm s}$ was lower in the HSS, implying a lower temperature of the plasma. Here we show that the higher $l_{\rm nth}/l_{\rm h}$ ratio of  the HSS is naturally explained with this synchrotron boiler model. 
However as will be discussed in Section~\ref{sec:discussion}  real spectral state transitions in X-ray binaries are a bit more complex and the spectral changes cannot be entirely attributed to a change in the soft photon compactness.

\subsection{Effects of $e$-$p$ Coulomb heating}\label{sec:effectsofCoulombheating}

\begin{figure*}
 \includegraphics[width=\textwidth]{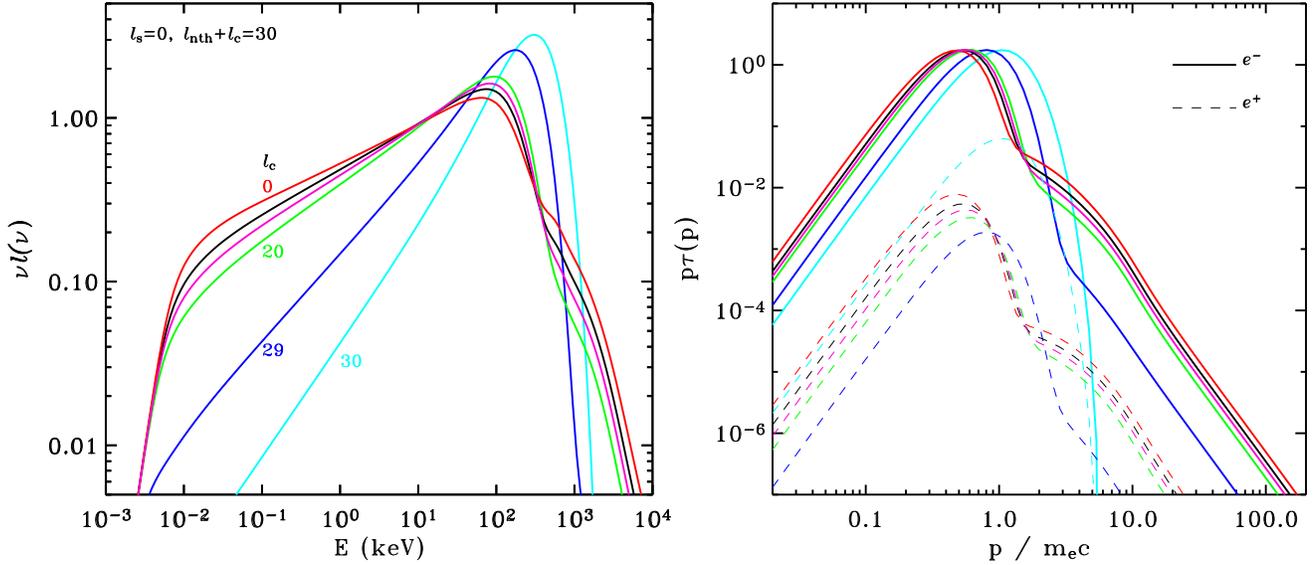}
 \caption{{ SSC models  with heating by hot protons.} Effects of varying the Coulomb heating compactness $l_{\rm c}$ at constant $l$ on the photon spectra  (left hand side) and the particle distributions (right hand side). The red curves show our fiducial model obtained for $l_{\rm c}=0$, $l_{\rm nth}=30$, $l_B=10$, $\Gamma_{\rm inj}=3$, $\gamma_{\rm min}=1$, $\gamma_{\rm max}=1000$, $\tau_{\rm i}=2$, $R=5 \times 10^7$ cm. The other curves show a sequence of models obtained  keeping $l=l_{\rm c}+l_{\rm nth}=30$ while the Coulomb heating compactness $l_{\rm c}$  is varied.  The black, pink, green, blue and cyan curves show the results for $l_{\rm c}$=10, 15, 20, 29, 30 respectively. In the right hand side panel, the full curves show the electron distributions while the dashed ones show the positron distributions.
 }
 \label{fig:lcoul}
\end{figure*}

In this section we investigate the effects of Coulomb heating by a thermal distribution of protons in a two-temperature plasma. Fig~\ref{fig:lcoul} shows the evolution of the spectrum when $l_{\rm c}$ is varied  at fixed $l=l_{\rm nth}+l_{\rm c}=30$,  $l_B=l/3$, $l_{\rm s}=0$. The other parameters are those of our fiducial model. When the fraction of Coulomb heating increases, the effects of the non-thermal component both in terms of high energy emission and soft cooling photon flux are reduced. As a consequence  the temperature of the thermal component increases. This leads to an increasingly  hard  spectrum peaking at increasingly large photon energies.  Finally, when Coulomb heating dominates, the equilibrium electron distribution is very close to a Maxwellian and, as a result, the non-thermal high energy tail is absent from the Comptonisation spectrum. 
Due to the effects of the pair production thermostat (Svensson 1984), the temperature of the electron plasma  is limited to  a few hundred keV. The temperature of the hot protons (shown in Table~\ref{tab:simus}) increases with $l_{\rm c}$ in accordance with equation~(\ref{eq:temp}).

\section{Application to Cygnus X-1}\label{sec:cygx}

In this section,  we  compare the results of our simulations to high energy spectra of Cygnus X-1. Since its discovery in 1964 (Bowyer et al. 1965),
this bright persistent source has been intensively observed by all the high energy instruments, from soft X-rays to $\gamma$-rays. It was the first dynamically proven black hole, and still remains the prototype of black hole candidates. 
The mass of the black hole is in the range 14--29 $M_{\odot}$ (Zi{\'o}{\l}kowski 2005) and its distance is 2.15$\pm$ 0.2 kpc (Massey, Johnson, 
\& Degioia-Eastwood 1995, see also the discussion and references in Zi{\'o}{\l}kowski 2005).

In most of the following we assume a black hole mass of 20 $M_{\odot}$ at a distance of 2 kpc.  Most of the X-ray emission is expected to be produced in a region of size $R\simeq 30 R_{\rm G}\simeq 10^8$~cm, where most of the accretion power is released. This is what we will assume in our simulations. However the observed luminosity may also arise from several distinct and more compact active regions (as in the patchy corona model for instance). In order to take into account the uncertainties in the source parameters and geometry, we denote:
\begin{equation}
\delta=\frac{30 R_{\rm G}}{nR}\frac{20M_{\odot}}{M}\frac{d^2}{d_{\rm o}^2},
\end{equation}
where $n$ is the number of distinct active regions in the accretion flow ($R$ is their typical size), and $d$ is the distance to the source and $d_{\rm o}$=2 kpc.
Then for a given observed flux $F$, the compactness parameter can be estimated as:
\begin{equation}
l=\frac{F}{F_o}\delta, 
\label{eq:fdelta}
\end{equation}
where $F_o$=6.8$\times$10$^{-9}$ erg s$^{-1}$ cm$^{-2}$.

\begin{table*}
 \centering
  \caption{Models parameters of the simulations shown in Fig.~\ref{fig:lcyg}. In addition to the models parameters the resulting proton temperature as well as the temperature of the thermal component of the electron distribution obtained from a fit with a Maxwellian are given. The last two columns give the ratio of magnetic to proton energy density (computed according to equation~(\ref{eq:beq})) and the magnetic to radiation energy density (computed using the radiation energy density resulting from the simulation).}
  \begin{tabular}{ccccccccccccccc}
  \hline
    Ref.         & $l_{\rm nth}$  & $l_{\rm c}$ & $l_{\rm s}$ & $l_{B}$ & $\tau_{\rm i}$  & $\Gamma_{\rm inj}$ & $\gamma_{\rm min}$ &        $\delta^{-1}$                  &$\tau_{\rm T}$ &  $kT_{\rm e}$ (keV) & $kT_{\rm i}$ (MeV) & $l_B/l_{B_{\rm P}}$ &  $l_B/l_{B_{\rm R}}$ \\
\hline
H1   &   2.51     &  28.1      &    1.8       &  0      &   1.34          &         2.00                &          1.5             &           0.13    &    1.45    &    97      &      13 & 0 &0\\ %042
H2   &  0.39       &  4.36      &    0.28    &   0     &   1.45          &         2.00               &           1.5             &          1                       &    1.45    &    92      & 1.6    &0 &0 \\%043
H3   &    1.00     &   3.00     &    0.29    &    0    & 1.45            &         3.00                &          1.5             &           1                         &   1.45    &     85      &   1.0 & 0 &0\\%041
H4    &    0           &   4.75     &   0           &    200     &   1.45    &                               &                             &            1                    &    1.45     &    83        &   1.3  & 37 & 131\\%064
H5    &     4.75     &  0           &    0          &   0.475   &   1.45    &         3.50               &             1             &           1                &    1.45    &    76        &     & 1.5 & 0.31\    \\%060

 \hline
S1     &  0.36       & 0.17     &    3.20     &      0      &    0.11       &         2.64              &          1.5             &       5.5    &   0.11      &    76                      &     9.4    &0&0\\%059
S2     &  1.98      & 0.96      &    17.6     &      0      &    0.11       &        2.64               &          1.5             &       1             &    0.12     &     67                     &     50 & 0 &0\\%198
S3     &   5.50     &   0          &    16.7     &      17.6 &    0.11      &        2.10               &            1               &            1          &    0.12     &    48                     &     &  1129  & 3.2 \\%049
 \hline
\end{tabular}
\label{tab:cyg}
\end{table*}

In order to test the synchrotron boiler model against spectra of accreting black holes it is important to have  good quality data extending above a few hundred keV and up to at least several MeV in order to estimate the flux level of the non-thermal tail.
With this regard, the average hard state spectrum of Cygnus X-1 obtained with the Compton Gamma-ray Observatory (\gro) probably represents the best data so far for a hard state source at MeV energies.  We use this \gro\ spectrum combined with a typical \sax\ hard state spectrum in order to extend the energy coverage down to the soft X-rays. Similarly we will also compare our results to the combined \sax/\gro\ spectrum  of the 1996 soft state of Cygnus X-1. These two spectra are shown on Fig.~\ref{fig:lcygdat}.  Both sets of data were  extensively presented and analysed in MC02. These authors fit both spectra with the hybrid thermal-non-thermal models {\sc eqpair} and found a very good agreement.   

However, as discussed earlier, the {\sc eqpair}  model does not take into account the effects of magnetic field  nor the Coulomb interaction with protons. In order to constrain the magnetic field and proton temperature we would like to fit those spectra with the results of our code. However our code has not been optimised yet and is too time-consuming in order to perform detailed spectral fitting. 
As a preliminary attempt to compare SSC models to spectra of accreting black holes, we will adopt a simpler approach and compare only qualitatively the shape of the simulated spectra with the unfolded data. Obviously, it will not be possible to compare the 'goodness' of different models nor evaluate the uncertainties on the model parameters.  Nevertheless  we will see that these simple 'fits by eyes'  already provide interesting constraints.
For both  spectra, we will proceed in three steps:
\begin{enumerate}
\item  First, we use the best fit {\sc eqpair} parameters found by MC02, replacing the unspecified thermal heating of {\sc eqpair} by Coulomb heating (i.e. setting $l_{\rm c}=l_{\rm th}$ and $l_B=0$), we check that our code is able to reproduce the shape of the spectrum, with a model without magnetic field. This allows us to estimate the proton temperature implied by the best fit models of MC02. This proton temperature depends on $l_{\rm c}$ (as can be approximated by equation~(\ref{eq:temp})). 

\item However, for a fixed (observed) X-ray flux,  $l_{\rm c}$ depends on the size and number of emitting regions as well as on the distance of the source ($l\propto \delta$).  In their spectral fitting procedure MC02 used an additional normalisation constant (directly related to $\delta$) which is a free parameter accounting for uncertainties on the size and distance and actually disconnecting the compactness parameter from observed luminosity. The best fit compactness parameters of MC02 lead to $\delta$ significantly different from unity. We therefore
compute a similar model with all compactness parameters divided by $\delta$ (so that the observed luminosity is produced for  $\delta=1$) and then check that we still have a good agreement with the data. If necessary, we modify the parameters to improve the fit and derive what we believe is a more realistic proton temperature. 

\item  We then set the Coulomb heating $l_{\rm c}=0$ and, allowing for the presence of the magnetic field, we explore the possibility of reproducing the spectrum with a pure non-thermal acceleration model. 
   
\end{enumerate}

For the purpose of our comparison with the data and in order to make our results comparable with the spectral fits of MC02, we now model the soft thermal emission of the cold accretion disc using a multicolour blackbody disc model  ({\sc diskbb} in XSPEC, Mitsuda et al. 1984) rather than a simple blackbody spectrum. When comparing our simulations with the observed spectra we account for the presence of a reflection component which is calculated using the {\sc pexriv} routine in XSPEC (Magdziarz \& Zdziarski 1995). We also take into account the standard column density, $N_{\rm H}=6\times10^{21}$ cm, of absorbing material in the direction of the source (we use the {\sc wabs} model in XSPEC).  The resulting model parameters are given in Table~\ref{tab:cyg} and the unabsorbed model spectra are displayed in Fig~\ref{fig:lcyg}.

\subsection{Low-hard state}\label{sec:lhscyg}

Cygnus X-1 spends most of the time in the LHS,  with a  rather stable flux of about 3.5$\times 10^{-8}$ erg s$^{-1}$ cm$^{-2}$. This sets a constraint on the compactness $l\simeq5\delta$ (see equation~(\ref{eq:fdelta})). 

Spectral fits with thermal Comptonisation models indicate that, in the LHS,  the optical depth is in the range $\tau_{\rm T}\sim1.5-3$ and the electron temperature 50 to 100 keV (DGK07). 
Then  equation~(\ref{eq:temp}) provides an upper limit on the proton temperature of about 0.1+1.9$\delta$ MeV.
This temperature would be achieved if the electron heating was fully due to Coulomb interaction.
If any other form of heating or acceleration is effective in Cygnus X-1 then the proton temperature must be lower. However, the observations show a non-thermal tail at high energy in excess of pure thermal Comptonisation model. While we have no direct evidence of Coulomb heating, the high energy excess indicates that non-thermal acceleration mechanisms are at work in the source. The temperature of the protons in the X-ray emitting region of Cygnus X-1 is therefore very likely to be much lower than typical ADAF temperatures (10--100 MeV). Our numerical simulations suggest even lower proton temperatures (see Table~\ref{tab:cyg}).
 
At equipartition with radiation, equation~(\ref{eq:eqprad}), gives $l_{B_{\rm R}} \simeq 2 \delta$ (i.e $B_{\rm R}\simeq 8 \times 10^5 \delta \sqrt{n}\ d_{\rm o}/d$ G) , and if equipartition is achieved with the protons then  $0.37<l_{B_{\rm P}}<0.44+7.9\delta$  depending on the  importance of Coulomb heating ($3.6 \times 10^5 \delta^{-1/2} < B_{\rm P}\delta^{-1}n^{-1/2} \ d/d_{\rm o}<1.7 \times 10^6$  G). 

\subsubsection{Best fit {\sc eqpair}  model}

For the hard state data, MC02 froze the disc temperature at 200 eV, and fixed $\gamma_{\rm min}=1.5$, $\gamma_{\rm max}$=10$^3$. Their best fit parameters are then $l_{\rm h}/l_{\rm s}=17^{+4}_{-3}$, $l_{\rm nth}/l_{\rm h}=0.082^{+0.088}_{-0.032}$, $\Gamma_{\rm inj}=2.0^{+0.9}_{-0.2}$, $\tau_{\rm i}=1.34^{+0.4}_{-0.5}$ and the soft compactness $l_{\rm s}=1.8^{+2.5}_{-1.6}$.  
They found a reflection component with a relative amplitude $\Omega/2\pi=0.52^{+0.06}_{-0.05}$ was required (reflection on neutral material was assumed).

\subsubsection{Models with $e$-$p$ Coulomb heating}\label{sec:mepcoulheat}

\begin{figure*}
 \includegraphics[width=\textwidth]{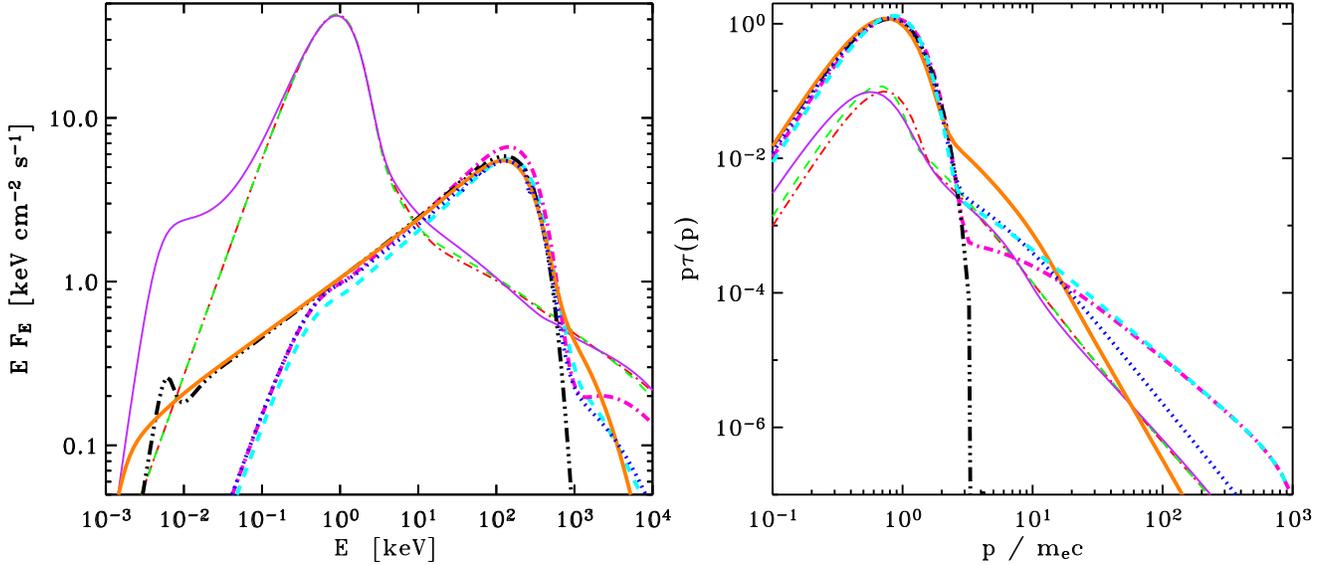}
 \caption{Various models providing a qualitative agreement with the observed high energy spectrum of Cygnus X-1, as described in Table~\ref{tab:cyg}.  The left hand-side panel shows the spectra while the right hand side panel shows the lepton distributions. The models for the soft state are shown by the thin curves (model S1: dot-dashed (red); S2: dashed (green);  S3: full  (purple)). The models for the hard state are shown by the thick curves (model H1: thick dashed (cyan); H2: dot-dashed (pink); H3:dotted (blue); H4: triple-dot-dashed (black); H5: full  (orange))} \label{fig:lcyg}
\end{figure*}

We used our code to simulate a similar model (referred as model H1 in Table~\ref{tab:cyg}). We used  the same parameters as those found by MC02 except that we  replaced the unspecified 'thermal' heating  compactness of {\sc eqpair} by Coulomb heating with $l_{\rm c}=l_{\rm th}$. The magnetic compactness was set to $l_{B}=0$ { and therefore the thermalisation occurs through the $e$-$e$ Coulomb boiler}. We found a spectrum which is very close to the {\sc eqpair} one, and that provides a good representation of the data. This spectrum is shown in Fig.~\ref{fig:lcyg}. As shown in Table~\ref{tab:cyg}, for this simulation (H1) the equilibrium proton temperature is  12.9 MeV. 

However, the compactness parameters of MC02 ($l\simeq$20), when compared to the observed flux, imply $\delta^{-1}\simeq0.13$, which we believe is unlikely. Our preferred  size and distance of the source (i.e. $\delta=1$) imposes a much lower value of the compactness ($l\simeq 5$) in order to reproduce the observed flux. If we adopt this value (model H2 in Table~\ref{tab:cyg}) we obtain a proton temperature of only 1.6 MeV. However, in this case the agreement with the \gro\ spectrum is not as good. The shape of the high energy excess is poorly reproduced.   

{ This is because the effect of $e$-$e$ Coulomb is more important  in the lower compactness model and flattens the slope of the non-thermal excess (as discussed in Section~\ref{sec:coulombboiler})}.
 Therefore, for a lower compactness an injection slope $\Gamma_{\rm inj}$ steeper than the one of MC02, is needed to fit the data. For instance, a better agreement could be found by setting $\Gamma_{\rm inj}=3$, $l_{\rm nth}=1$, $l_{\rm c}=3$, which leads to a proton temperature of 1 MeV (model H3 in Table~\ref{tab:cyg}) . The photon spectra corresponding to these three models can be compared in Fig.~\ref{fig:lcyg}.

For comparison, Fig.~\ref{fig:lcyg} also shows a pure thermal SSC model ($l_{\rm s}=0$,     
 $l_{\rm nth}=0$, model H4 in Table~\ref{tab:cyg}) leading to a proton temperature of about 1.26 MeV. The agreement with the data requires an excessively super equipartition magnetic field ($l_B/l_{B_{\rm P}}$$\sim$30) and of course does not reproduces the non-thermal high energy excess.  
 This confirm the results of Wardzi{\'n}ski \& Zdziarski (2000 here after WZ00) who derived analytical estimates of the thermal SSC luminosity, compared them to the luminosity of several black hole sources and showed that the magnetic field would have to be implausibly  strong in order to match the observed luminosity. For Cyg X-1 however,  they  
found that for equipartition magnetic field and typical parameters, thermal SSC represents a significant fraction, and could possibly dominate the observed luminosity. 
This is not in contradiction with our results.
In fact, these authors estimate the same magnetic field as suggested by our results, but they assume a proton temperature ten times larger (10 MeV) and an emitting region 3 times smaller, so that they find a proton energy density 30 times larger. Although in this case, the magnetic field is indeed in equipartition with matter, we have shown that the proton temperature must be smaller, which implies a magnetic field clearly above equipartition.

\subsubsection{Model with pure non-thermal acceleration}

Alternatively, the hard state spectrum can be quite well represented with a pure SSC model  which does not include any form of thermal or Coulomb heating. Fig~\ref{fig:lcygdat} shows a comparison of the observed spectrum with a simulated spectrum with  $l_{B}=0.475$, $l_{\rm nth}=4.75$, $l_{\rm c}=0$, $l_{\rm s}=0$, $\Gamma_{\rm inj}=3.5$ (model H5 in Table~\ref{tab:cyg}). A steep power-law electron injection slope is required in order to reproduce the hard slope of the hard X-ray spectrum (for flatter injection, the soft cooling synchrotron flux is too strong). It also helps fitting the steep slope of the non-thermal excess. The magnetic field is approximately at equipartition with the particles and about 30 percent of equipartition with the radiation field.  Note that in this model we set the minimum injection Lorentz factor of the electron to 1 (while MC02 use 1.5). Again this has very weak  effects on the resulting spectrum.

 { We also note that in this model the radiation energy density dominates over that of the gas, which might be a cause of instability. This can be avoided if the size of the emitting region is larger than what we assumed by at least a factor of 5 (e.g. 150 $R_{\rm G}$ instead of 30 $R_{\rm G}$).}
{ Finally, it is apparent in Fig.~\ref{fig:lcygdat} that our model underestimates the flux below 1 keV.  This may be due to the fact that our model does not include any disc thermal component.  Such a thermal component in the LHS  was firmly established by previous studies.  However, as discussed in Section~\ref{sec:softphotoncompactness}, it is possible  that this thermal disc photon flux does not intercept the X-ray emitting region and has no effect on the cooling of the coronal electrons. }

\subsection{High soft state}

In the HSS, the source is about 4 times more luminous than in the hard state (see MC02). This leads to $l\simeq20\delta$. At equipartition with radiation the magnetic field compactness is $l_{B_{\rm R}}\simeq5\delta$ (or $B_{\rm R}\simeq 1.5 \times 10^6 \delta \sqrt{n}\  d_{\rm o}/d$ G).

\subsubsection{Best fit {\sc eqpair} model}

The soft state spectrum was fit by MC02 with {\sc eqpair}  with a disc temperature of 370 eV, and the best fit parameters were $l_{\rm h}/l_{\rm s}=0.17^{+0.01}_{-0.01}$, $l_{\rm nth}/l_{\rm h}=0.68^{+0.20}_{-0.12}$, $\Gamma_{\rm inj}=2.6^{+0.2}_{-0.2}$, $\tau_{\rm i}=0.11^{+0.02}_{-0.02}$, and a ionised reflector (with ionisation parameter $\xi=100^{+210}_{-60}$ and reflection amplitude $\Omega/2\pi=1.3^{+0.3}_{-0.3}$). 
Because , in their fit,  compactness is decoupled from the observed luminosity (i.e. $\delta$ is a free parameter), they find the soft photon compactness is poorly constrained. They find $l_{\rm s}=3.2^{+38}_{-0.12}$.

\subsubsection{Models with Coulomb heating}

We simulated a similar model replacing $l_{\rm th}$ by $l_{\rm c}$ (model S1 in Table~\ref{tab:cyg}) and again we found a spectrum that was very similar to that of {\sc eqpair} and consistent with the \sax/\gro\ data.
For this set of parameters we derive a proton temperature of 9.35 MeV.  
The compactness parameter then  implies $\delta^{-1}$=5.5.
We note that according to the fit of MC02,  $\delta^{-1}$ in the soft state is larger than in the hard state by a factor of 42.  This would imply that $nR$ is larger in the soft state, which appears in conflict with the timing data suggesting the size of the emitting region is smaller in the HSS (see e.g. DGK07). 
For our assumed size of the emitting region of 10$^8$ cm and a distance of 2 kpc (i.e. $\delta\simeq1$) , we need a soft photon compactness  $l_{\rm s}=17.6$ to match the observed flux. 
The resulting proton temperature is then 50 MeV (see model S2 in Table~\ref{tab:cyg}).

\begin{figure}
 \includegraphics[width=\columnwidth]{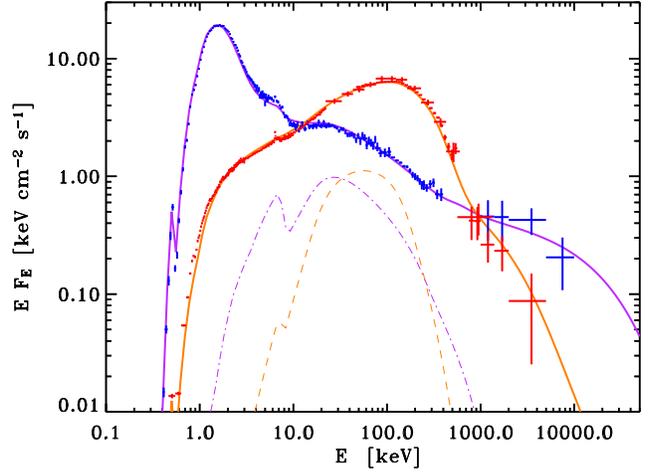}
 \caption{A comparison of the average \gro\ spectra for the HSS (blue)  and  LHS (red) of Cygnus X-1 from MC02, with models involving only injection of non-thermal particles as heating mechanism. At low energy the \gro\ data are complemented by \sax\ data (see MC02 for details). 
 The thick purple line shows our absorbed soft state model S3 which parameters are given in Table~\ref{tab:cyg}, while the orange line shows the hard state model H5 of the same table. Reflection components were added to both spectra and are shown by the thin dot-dashed and dashed curves for the soft state model and the hard state model respectively.}
 \label{fig:lcygdat}
\end{figure}

\subsubsection{Model with pure non-thermal acceleration}\label{sec:hssnonther}

Alternatively, a qualitative agreement with the average soft state spectrum of MC02 can be obtained without any Coulomb heating ($l_{\rm c}=0$). Fig~\ref{fig:lcygdat} shows a comparison of the data with a simulated spectrum with $l_{B}=17.6$, $l_{\rm s}=16.7$, $l_{\rm nth}=5.5$, $\Gamma_{\rm inj}=2.1$ (model S3 in Table~\ref{tab:cyg}). 
This suggests a relatively strong magnetic field of 2.3 $\times$ 10$^{6}$ G (about 3 times equipartition with radiation). Lower magnetic fields do not give a satisfactory description of the observed spectrum because the thermal component of the electron distribution is then too cool (the shape of the spectrum in the 2-20 keV band is not reproduced). A higher magnetic field seems required in order to heat up the plasma (in the way  discussed in Section~\ref{sec:ls}).

Our results for the soft state show that the spectrum can be reproduced either assuming cold protons and a magnetic field above equipartition, or no magnetic field and a relatively high temperature of the protons. The magnetic and the proton thermal energy density of the two models are comparable. Obviously intermediate models with both magnetic field and hot protons are not excluded and would allow for a lower proton temperature and/or magnetic field.

\section{Discussion}\label{sec:discussion}

The main result of this paper is that both the LHS and and the HSS spectra can be produced with pure non-thermal acceleration models with no need for thermal heating (as assumed in the {\sc eqpair} model) or hot protons (as assumed in ADAF-like models). 
In the framework of accretion disc corona model, this explains how  the electron distribution could look so different in both spectral states while the energy dissipation processes remain the same (i.e. non-thermal acceleration). In the context of the hot accretion flow models for the hard state, our result explains why the electron distribution is essentially thermal even if, as expected,  non-thermal acceleration/heating processes are important (as also shown by Mahadevan \& Quataert 1997).
In the following we summarise and discuss further the constraints on the main parameters of the model.

\subsection{Soft photon compactness}\label{sec:softphotoncompactness}

\subsubsection{Low hard state}

In the LHS of several sources, it is likely that Compton scattering off soft disc photons  contributes to the high energy emission and could even dominates over SSC. This is suggested by the correlation that is observed in Cygnus X-1 and other sources  between the strength of the disc reflection features and the spectral slope of the X-ray emission, indicating a stronger Compton cooling when the disc is 'closer' to the X-ray emitting region (Zdziarski, Lubi{\'n}ski \& Smith 1999; Gilfanov,
Churazov \& Revnivtsev 1999).  Such disc thermal emission is also clearly detected in the LHS. However it  is possible that it does not  intercept the Comptonising cloud. This would be the case if  this emission arises from a disc truncated at large distances from the central region where most of the Comptonised emission is produced. In any case our results demonstrate that such external soft seed photons are not required by the data. The broadband high energy spectrum of Cygnus X-1 appears to be consistent with a pure SSC model with a slightly sub-equipartition magnetic field.  Our results, of course,  do not exclude that  Comptonisation of external soft photons could dominate in the LHS. This would simply require either a lower magnetic field, or some heating by hot protons in order to keep the electron temperature around 100 keV.  

\subsubsection{High soft state}

We have  shown that in the HSS the presence of a strong illumination by soft photons has important consequences on the coronal electron distribution. This soft photon flux cools down the thermal particles so that most of the hard X-ray luminosity is produced by the non-thermal electrons. 

\subsubsection{Spectral transition}

The main parameter controlling the spectral evolution during state transitions is the soft photon compactness.  
The change from quasi-thermal  to essentially non-thermal Comptonisation spectrum is achieved naturally once the effects of synchrotron boiler are taken into account. 
However, the data indicate that other coronal parameters change during spectral state transitions. Broad band spectral fits of HSS spectra of Cygnus X-1 with {\sc eqpair} by  G99 and  MC02  suggest the Thomson optical depths is lower by at least a factor of 5 in the HSS. The situation is therefore more complex. Our comparison with the average soft and hard state spectra of Cygnus X-1 (presented in Section~\ref{sec:cygx}) also shows that other coronal parameters must be changed in order to reproduce the details of the observed spectra.

\subsection{Magnetic field}\label{sec:b}

It is interesting to express the magnetic field values we found in our comparison with the data as a fraction of the  equipartition compactness (radiation or particles). Indeed, because the shape of the spectrum depends on the ratio $l_B/l$ rather than on the absolute value  $l_B$, the ratios and  $l_B/l_{B_{\rm R}}$ is independent of the uncertainties on the source size and distance (i.e. $\delta$) and $l_B/l_{B_{\rm P}}$  depends on $\delta$ only at low proton temperature. 

\subsubsection{Low hard state}

We qualitatively reproduce  the LHS spectrum of Cygnus X-1 with a pure non-thermal model  in which the magnetic field is close to equipartition  with the particle energy ($l_B \simeq 3y/8)$. We note that Wardzi{\'n}ski  and Zdziarski (2001) (hereafter WZ01) using analytical estimates for the SSC luminosity accounting of the synchrotron  emission of  the non-thermal  electrons,  found that the magnetic field in Cyg X-1 must be strongly sub-equipartition with the protons\footnote{On the contrary WZ00 assumed the electron distribution is a pure Maxwellian. They found that  the magnetic field would then be large, which is in agreement with our result for a pure Coulomb heating model, as discussed in section~\ref{sec:mepcoulheat}}. This is because they assumed a much higher proton temperature of 10 MeV, while in our model without Coulomb heating (H3)  the  protons have the same temperature as the electrons (80 keV). More generally the observed luminosity sets an upper limit to the proton temperature of about 2$\delta$ MeV as shown above. Actually, our constraints are fully consistent with  those of WZ01.
In any case, the magnetic field infered from model H3 is probably an upper limit on the actual magnetic field in the source. The presence of external soft photons would imply a lower $B$ as mentioned just above. If Coulomb heating is important then $l_{\rm nth}$ must be lower to conserve  the same luminosity. At the same time, in order to keep the same non-thermal Compton luminosity it would be necessary to reduce the synchrotron losses and probably decrease $B$, and then the $l_B/l_{B_{\rm P}}$ would be lower also due to the higher temperature of the protons. 

On the other hand, this maximum magnetic field  appears significantly below equipartition with radiation. Our comparison with the data of Cygnus X-1 suggests, $l_B/l_{B_{\rm R}}<$0.3. 
 Nevertheless it suggests that the emission of the corona is not powered by the magnetic field, as assumed in most accretion disc corona models (see e.g.  di Matteo, Celotti \&  Fabian 1997; Merloni \& Fabian 2001).
{  Indeed if the corona is powered by the magnetic field, then the luminosity can be written as $L=VU_{B}/t_{\rm dis}$ where $t_{\rm dis}$ is the magnetic field dissipation time scale.  
Since $L=U_{\rm R}V/t_{esc}$, we have $U_{B}/U_{\rm R}=t_{\rm dis}/t_{\rm esc}$. Our constraint on the magnetic field in Cygnus X-1 then translates into:
\begin{equation}
t_{\rm dis}\la0.3(1+\tau_{\rm T}/3)R/c\simeq0.45R/c\label{eq:tdis}
\end{equation} 
 which points toward extremely fast magnetic dissipation. The magnetic dissipation time-scale is expected to be $t_{\rm dis}\sim R/v_{\rm A}$ with $v_{\rm A}<c/3$ is the Alfven speed.  This would gives $t_{\rm dis}\ga3R/c$, in contradiction with equation~\ref{eq:tdis}.}

{ We stress however that this
requirement of a low magnetic field in the LHS is entirely due to  the observed strength of the non-thermal  MeV excess in the spectrum of Cygnus X-1  and our interpretation of it as the non-thermal tail of the Comptonising electron distribution.  If contrary to our one-zone hypothesis, the  MeV excess is produced in a different region than the bulk of the thermal Compton emission, the constraint might be  relaxed.} 

\subsubsection{High soft state}

In the HSS the magnetic field (2.3 $\times$ $10^6$ G for our assumed size of the emitting region) is 6 times larger than in the LHS, with a magnetic energy density exceeding both the particle and radiation energy density. The magnetic field must be higher in the HSS, in order to keep the temperature of the thermalised electrons high enough (see Section~\ref{sec:hssnonther}). Such a magnetic field would be strong enough to power the corona and this supports the magnetic corona model for the HSS.  We note however that spectral fits of other HSS spectra of Cygnus X-1  by G99, do not require any heating of the thermal electrons ($l_{\rm th}=0$), in which case the magnetic field would be much lower. Actually, constraining the magnetic field in the HSS would require a more detailed comparison with a larger set of data.

\subsection{Coulomb heating by hot protons}\label{sec:coulombheating}

{ Our model including Coulomb heating  by hot protons provides a good description of the spectra of Cygnus X-1, although in both states Coulomb heating is not enough and some level of non-thermal acceleration is required to explain the non-thermal tails.}

\subsubsection{Low hard state of Cygnus X-1}
 
 In the LHS, both our simulations and analytical estimates suggest a proton temperature  which is rather low when compared to what obtained in typical two-temperature solutions. The low proton temperature we infer is related to the measured Thomson optical depth and luminosity. The optical depth in Cygnus X-1 is larger than unity and therefore the electrons are quite efficiently coupled with the protons. Hot protons would produce a luminosity that is larger than observed.
We estimated an upper limit on the proton temperature at about 2$\delta$ MeV, possibly much lower if there are other forms of electron heating or acceleration. In contrast, typical ADAF solutions have $kT_{\rm i}$ in the range 10--100 MeV.  Having a temperature comparable to ADAF models would require a size of the emission region more than 5 times smaller than the value ($10^8$ cm)  assumed in this work. In Cygnus X-1, the hard state emission is well modelled as coming from a hot inner flow surrounded by a cold disc truncated at  30--100 $R_{\rm G}$, (e.g., DGK07). If so, having a high proton temperature would imply a mass of the black hole lower than $4 M_{\odot}$ which is clearly excluded.

 \subsubsection{Low hard states of XTE~J1118+480 and GX~339--4}
We cannot exclude however that a combination of low mass, smaller size (in terms of gravitational radii) and uncertainties related to our assumption of an homogeneous spherical region, may combine to underestimate the proton temperature in Cygnus X-1 by a factor of 5.  
However, there are  other sources which have a much lower luminosity and similar Comptonisation parameters. For instance Frontera et al. (2001b) analyse the \sax spectrum of XTE J1118+480  during its outburst of 2000, while the source had a luminosity of about one order of magnitude lower than Cygnus X-1.  The observed X-ray flux of XTE J1118+480 was 4.8$\times 10^9$ erg cm$^{-2}$ s$^{-1}$, which injected  in equation~(\ref{eq:fdelta}) and~(\ref{eq:temp}) gives $T_{\rm i}/T_{\rm e}<1+2\delta$. Fixing the black hole mass at its minimum possible value (6 $M_{\odot}$) and largest distance allowed by the uncertainties (d=1.8$\pm0.6$ kpc, McClintock et al. 2001), and keeping $R$=30 $R_{\rm G}$, one obtain a conservative upper limit on  $T_{\rm i}/T_{\rm e}<$10 which is much lower than what is expected in two-temperature accretion flows. Then reaching high protons temperatures would require a very small $R<3R_{\rm G}$. Similar conclusions could be reached for the source GX~339--4 which was observed at comparable flux levels, with similar electron temperature and optical depth (see e.g. Joinet et al. 2007). 
For lower luminosity sources, it becomes difficult to obtain robust constraints on the optical depth.  At least, for sources above 10$^{-3}L_E$ a two-temperature plasma seems quite unlikely.

\subsubsection{Comparison with standard two-temperature accretion flow models}

Yuan \& Zdziarski (2004) show that two-temperature accretion flow  models tend to produce higher electron temperatures and lower optical depths than what is typically measured in Cygnus X-1, GX~339--4 or XTE~J1118+480. This is in agreement with what we find here:  a high proton temperature would indeed require a Thomson optical depth lower than what is observed. Also among the simulations shown in Yuan and Zdziarski (2004), the models in which the plasma parameters are the closest to the observed ones have a luminosity much larger than that of GX~339--4 and XTE J1118+480 in their low luminosity states.

Esin et al. (1998) presents a comparison of the ADAF model spectra with  {\it Ginga} and \gro/OSSE  data of Cygnus X-1 in the LHS. In this comparison, Esin et al. assume a black hole mass of  $M$=9$M_{\odot}$ and a distance d=2.5 kpc, which  (for $R$=30 $R_{\rm G}$) implies $\delta$ =3.5.
They  found a relatively good agreement for the shape of the spectrum. However, although they assumed a large $\delta$, their model luminosity is still a factor of three larger than what is observed (they had to re-normalise their model spectra to fit the observations). This is in agreement with the estimates presented here: with $\delta=3.5$ a luminosity larger by a factor of 3 would allow for a proton temperature larger than 10 MeV.

Esin et al. (2001) show that the simultaneous {\it Chandra}/\xte\ spectra of XTE J1118+480
is qualitatively matched by the ADAF model. Unfortunately the ion temperature and the plasma parameters are not given in this paper and we cannot understand this result. However we note that the shape of the model spectra  present significant differences  with the data at high energy, which suggests that the temperature and optical depth of the ADAF are different from the accurate value measured  by Frontera et al. (2001) with \sax. Similarly, Yuan, Cui \& Narayan (2005) compare the same data with a coupled ADAF/jet model (in which the X-ray emission is dominated by the ADAF) and find a good agreement. { Nonetheless, in their best fit model, the region where most of the radiation is produced  has a Thomson optical depth $\tau_{\rm T}$ of only 0.2 and the electron temperature is  around  600 keV (F. Yuan private communication)}. Unlike the \sax\  data, the \xte\  data do not allow one to constrain the electron temperature. Because of their lack of sensitivity around 100 keV, the \xte\ data allow for optically thin high temperature solutions that are naturally produced in ADAF models but would probably be excluded by the \sax data. 

In any case, our estimates for the proton temperature also relies on the poorly constrained  size of the emitting region and we can not completely exclude that this dimension is much smaller than 30$R_{\rm G}$. Actually, the emissivity of an ADAF and of other two-temperature accretion flows depends on the distance to the black hole (or radius). 
Our size of the emitting region $R$ should correspond to the radius (or range of radii) where most of the luminosity is radiated.  
Yuan \& Zdziardski (2004) show that, at least at high luminosity,  the emission rate per logarithm of radius peaks at about 3--4 $R_{\rm G}$. This suggests a much more compact emitting region than the one considered here. This could allow such models to match the spectra and luminosity of Cygnus X-1 and possibly XTE~J1118+480.  An investigation of whether and how this is possible, would be very interesting but is largely out of the scope of the present work. 

 We therefore conclude that in the HSS of  black hole binaries with accurately measured optical depth ($\tau_{\rm T}\simeq$1.5--3)  and electron temperature ($kT_{\rm e}\simeq$50--100 keV), a hot two-temperature accretion flow is viable only if the X-ray emission is concentrated in a very  compact region (lower than a few $R_{\rm G}$).  

\subsubsection{High soft state}

In contrast in the HSS of Cygnus X-1 the optical depth is lower  than in the LHS, and a much higher proton temperature ($\sim$50 MeV) is allowed by the energy balance. This would be in agreement  with two-temperature accretion disc coronae models that have been proposed as alternative to magnetically dominated coronae (di Matteo, Blackman \& Fabian 1997; R{\'o}{\.z}a{\'n}ska 
\& Czerny 2000).  However, from the power-law shape of the high energy HSS spectra most of the power must be released to the electrons through non-thermal processes.
Moreover, as mentioned in Section~\ref{sec:b}, it is not clear whether thermal heating is really required in the HSS and the protons could be cold as well.  

\subsection{Slope of the injected electron distribution and acceleration processes}

\subsubsection{Constraints on the injection slope of the electrons}

 In the HSS, the slope of the injected electrons $\Gamma_{\rm inj}$ is directly constrained by the shape of the high energy powerlaw spectrum (in the Thomson limit, the photon index $\Gamma=\Gamma_{\rm inj}/2+1$).  Our rough comparison allows us to constrain $\Gamma_{\rm inj}$ in the range  2.1--2.7 depending on the details of the model. In the LHS on the other hand the slope of the non-thermal electron distribution is not very well constrained by the shape of the weak high energy excess. However, if injection of non-thermal particles is the only form of heating for the plasma, then this slope must be much  steeper ($\Gamma_{\rm inj} > 3.5$) in order to limit the synchrotron emission (which otherwise would cool down the thermal component). Even if some form of thermal heating of the electrons (through Coulomb heating for instance) can maintain the high temperature of the electrons, the injection slope must be steep in order to keep the SSC luminosity at the observed level. This is a very strong constraint on the injection mechanism in the LHS.

 \subsubsection{Consequences for acceleration models}
 
 The different injection slopes in the HSS and LHS could be due to different acceleration mechanisms, but they could also be due to the same physical acceleration process operating in a different regime (e.g. different magnetic and photon fields,  electron density...).  Investigating this issue would require a more detailed modelling of the acceleration processes. Our results indicate that neither  magnetic dissipation nor Coulomb heating in a standard two-temperature plasma, is the main heating/acceleration  process of the corona in the hard state.
 
There is also another important issue regarding the  acceleration process: It is not clear whether injecting non-thermal electrons with a power-law distribution provides a good description of the acceleration processes at work in the corona of black hole binaries. For instance, Belmont et al. (2008b) show that when the synchrotron boiler effects are taken into account, stochastic acceleration processes in a magnetised plasma lead to an electron distribution which is very close to a Maxwellian. Departure from a Maxwellian occurs only  in some specific conditions when the stochastic acceleration process can accelerate particles only above a given threshold energy. A more detailed modelling and investigation of these acceleration processes and their relevance to the observations is deferred to a future work. However, the results of Belmont et al. (2008b),  and, in a different context, those of  Katarzy\'nski et al. (2006), already show that it is difficult to avoid a large fraction of the power being transferred directly to the thermal electrons. If non-thermal acceleration processes indeed lead to a direct thermal heating of the electrons, this would reduce room for a contribution from Coulomb heating in the LHS.  For instance, this would certainly reduce dramatically the uppers limits on the proton temperature that we have set for Cygnus X-1 and possibly exclude the presence of a two-temperature plasma in this source. On the other hand modelling the high soft state with such  acceleration models might be more challenging and should set serious constraints on the acceleration mechanisms. 
\section{Conclusion}

We have explored the effects of the synchrotron self-absorption on the particle energy distribution and emitted radiation of the corona of accreting black holes. We have found that, in both spectral states of black hole binaries the 
 coronal emission  could be powered by a similar non-thermal acceleration mechanism. In the LHS the synchrotron and $e$-$e$ Coulomb boilers redistribute the energy of the non-thermal particles to form and keep a quasi-thermal electron distribution at a relatively high temperature, so that most of the luminosity is released through quasi thermal comptonisation. In the HSS,  the soft photon flux from the accretion disc becomes very strong and cools down the electrons, reducing the thermal Compton emissivity. Then most of the emission is produced by disc photons up-scattered by the non-thermal cooling electrons. Another difference between the two states is that the slope of the injected electrons has to be steeper in the LHS to reduce the synchrotron emission.   
 
Our comparison of simulations with the high energy spectra of Cygnus X-1 in the LHS allowed us to set upper limits on the magnetic field and the proton temperature.  Our results indicates that the  magnetic field is at most at equipartition with the particle energy density and below equipatition with radiation (unlike what is assumed in most accretion disc corona models). The proton temperature is found to be lower than 2$\delta$ MeV. 

{ The constraint on the proton temperature is independent of the synchrotron boiler effects (it is valid even if the magnetic field B=0)}.  Indeed, in a two-temperature plasma, due to electron ion coupling, there is a relationship between the luminosity, the size of the emitting region,  the ion and electron temperatures, and the Thomson optical depth. For the measured luminosity, electron temperature and optical depth of several black hole binaries, like XTE~J1118+480 or GX~339--4, a proton temperature much higher than the electron temperature $T_{\rm i}/T_{\rm e}\sim$100--1000 (as in a two-temperature accretion flow) requires an extremely small size of the emitting region (a few $R_{\rm G} $ or less).

 In the HSS the constraints on the magnetic field and the proton temperature are less stringent. The magnetic field could  power the emission of the corona. 

However these constraints on the plasma parameters rely on qualitative comparisons of the simulated spectra with the data. It would be interesting to refine and set more quantitative constraints on the magnetic field, the proton temperature and other physical parameters of the corona of accreting black holes. We hope to present  such detailed spectral fits in a forthcoming work. 

\section*{Acknowledgments}
We thank Andrzej Zdziarski and Andrea Merloni for useful comments and suggestions, we also thank Feng Yuan for discussions on ADAF models and for providing us with unpublished data.  
JM is grateful to the Institute of Astronomy in Cambridge (UK) for hospitality during the final stages of this work. This research was funded by CNRS and ANR.

\newpage
\appendix
\section{Analytical estimates of the steady-state non-thermal electron distribution and Synchrotron turn-over frequency}\label{sec:appendix}

The Cyclotron frequency,
 \begin{equation}
 \nu_{\rm c}=\frac{eB}{2\pi m_{\rm e} c},
 \end{equation}
 can be expressed in terms of the parameters of our model as follows:
 \begin{equation}
 \epsilon_{\rm c}=h\nu_{\rm c}/m_{\rm e}c^2=\sqrt{\frac{3\lambda l_{\rm B}}{4\pi\alpha_{\rm s} R}},
 \label{eq:ec}
 \end{equation}
   where $\lambda$ is the Compton wavelength of the electron and $\alpha_{\rm s}$ is the fine structure constant. 
   
  For a relativistic electron of Lorentz factor $\gamma$ most of the synchrotron power is emitted around $\nu\simeq2\gamma^2\nu_{\rm c}$. Therefore, for a wide range of electron energy distributions, the synchrotron emission at a frequency $\nu$ is dominated by electrons of Lorentz factor:
 \begin{equation}
 \gamma\simeq\sqrt{\frac{\epsilon}{2\epsilon_{\rm c}}}=15.68\sqrt{\frac{\epsilon\times 511}{10^{-2}}}\left(\frac{R}{5\times 10^7 {\rm cm}}\frac{10}{l_{\rm B}}\right)^{1/4}
 \end{equation}
where $\epsilon=h\nu/m_{\rm e}c^2$.

 For non-thermal electrons the radiative losses  can be written as follows:
 \begin{equation}
 \dot\gamma_{r}=-\frac{4c}{3R}\gamma^2\beta^2 l_{\rm B}\left[1+g\right],
 \end{equation}
  where $g$ accounts for the Compton losses. 
It is easy to see that  $g$ is larger at large $l/l_{B}$. It is however rather difficult to estimate accurately. Because most of the radiation energy density is in the form of X-ray photons, the Compton interactions with the non-thermal electron occur mainly in the Klein-Nishina regime. As a consequence $g$ is considerably reduced with respect to its expected Thomson limit value. For the models with $l_{\rm nth}/l_{\rm B}=3$, the radiation and magnetic field  energy density are comparable and yet the losses of the non-thermal particles are dominated by synchrotron ($g\simeq0.1$). Another complication is that in the Klein-Nishina limit  $g$ depends slightly on $\gamma$ (i.e. the Compton losses do not scale linearly with $\gamma^2\beta^2$). Since in most of our models, synchrotron dominate the losses of the non-thermal electrons, we can neglect this dependence and a rough estimate of  $g$ is enough. 

We found that a reasonable approximation of $g$ could be obtained  by neglecting the losses on the Comptonised X-ray photons. If one consider as targets only  the synchrotron and  external soft photons in the Thomson regime:
 \begin{equation}
 g\simeq\frac{3}{4\pi}\frac{l_{\rm sync}+l_{\rm s}}{l_{\rm B}}\left(1+\frac{\tau_{\rm T}}{3}\right),
 \label{eq:g}
 \end{equation}
where 
\begin{equation}
l_{sync}=\frac{L_{sync}\sigma_T}{Rm_{\rm e}c^3},
\label{eq:lsydef}
\end{equation}
and $L_{\rm sync}$ is the self-absorbed synchrotron luminosity of the electrons. 
An estimate for $l_{\rm sync}$ is derived below which then allows to evaluate $g$  (see equations~(\ref{eq:lsync}) and (\ref{eq:geq})).

 On the other hand the $e$-$e$ Coulomb losses can be estimated as:
\begin{equation}
\dot\gamma_{c}\simeq-\frac{3c}{2R}\frac{\ln\Lambda\tau_{\rm T}}{\beta\langle\gamma\rangle}, 
\end{equation} 
 where $\langle\gamma\rangle$ is the average Lorentz factor of the  targets electrons (see Nayakshin \& Melia 1998) and is close to unity (we use $\langle\gamma\rangle$=1.15 in our numerical evaluations).
And therefore:
\begin{equation}
\frac{\dot\gamma_{\rm r}}{\dot\gamma_{\rm c}}=\frac{8}{9}\frac{l_{\rm B} (1+g)\langle\gamma\rangle}{\ln\Lambda \tau_{\rm T}}\beta^3\gamma^2.
 \end{equation}
 At high energies, the losses are dominated by Compton and synchrotron while at low energies Coulomb dominates the cooling. The transition from ratiative to Coulomb cooling occurs at a reduced  momentum $p_{\rm c}$ such that: 
 \begin{equation}
 p_{\rm c}^6=\xi^2(p_{\rm c}^2+1)\label{eq:pc},
 \end{equation}
 where 
 \begin{equation}
 \xi=\frac{9}{8}\frac{\ln\Lambda \tau_{\rm T}}{l_{\rm B} (1+g)\langle\gamma\rangle}.
 \end{equation}
 
 At large $\xi$, $p_{\rm c}\simeq\sqrt\xi$ while at small $\xi$, $p_{\rm c}\simeq\xi^{1/3}$.
 For the cases, $l_{\rm B}$=0.1, 10, and 1000, this gives $p_{\rm c}$=19.4, 1.98 and 0.56 respectively. 
 $\xi$ characterises the relative importance of $e$-$e$ coulomb as a cooling process.  Coulomb collisions become less and less important  at  larger compactness. 
 
 Using the `stationary' approximation (see e.g. Fabian et al. 1986, Ligthman \& Zdziarski 1987, Coppi 1992), the steady state non-thermal electron distribution can be estimated as:
 \begin{equation}
 \tau(\gamma)=-\int_{\gamma}^{\gamma_{\rm max}}d\gamma\quad \dot{\tau}(\gamma)
 /(\dot{\gamma_{\rm r}}+\dot{\gamma_{\rm c}}),
 \end{equation}
 where $\dot{\tau}(\gamma)$ is the distribution of injected electrons.  
 Assuming  $\gamma_{\rm min}=1$ and $\Gamma_{\rm inj}>2$,  for $1\ll\gamma\ll\gamma_{\rm max}$, the non-thermal particle distribution can be estimated as 
 \begin{equation}
 \tau(\gamma)=\tau_0\gamma^{1-\Gamma_{\rm inj}}p^{-2}/(1+\xi\gamma p^{-3}),
 \label{eq:analdistrib}
 \end{equation}
with 
 \begin{equation}
 \tau_0=\frac{9}{16\pi}\frac{l_{\rm nth}}{l_{\rm B}}(1+g)^{-1}F(\Gamma_{\rm inj}),
 \end{equation}
 and
 \begin{equation}
 F(\Gamma_{\rm inj})=\left(\frac{\Gamma_{\rm inj}-1}{\Gamma_{\rm inj}-2}-\langle\gamma\rangle\right)^{-1}.
 \end{equation}
 This expression is in good agreement with the results of our simulations  for $\gamma> \gamma_{\rm t}$ and when the Compton losses are not too strong (i.e. $l_{\rm nth}/l_{B}\la10$). At lower energies synchrotron self-absorption effects become important and the stationary approximation gradually breaks down.  
 
 At energies above $p_{\rm c}$, the distribution is a powerlaw  $\tau(\gamma)=\tau_0\gamma^{-s}$ with $s=1+\Gamma_{\rm inj}$.
For such a distribution the reduced synchrotron self-absorption frequency $\epsilon_{\rm t}$ is given by (see WZ01):
 
 \begin{equation}
 \epsilon_t=\left(\frac{3^{(3+s)}}{2^8}\frac{\pi}{\alpha_{\rm s}}\tau_0^2 G_1^2\right)^{1/(4+s)}\epsilon_c^{(2+s)/(4+s)},
 \label{eq:et1}
 \end{equation}
 where $G_1\simeq1$  is defined in equation (9) of WZ01.
 
 Then the Lorentz factor of the electrons emitting around the turn over frequency, $\gamma_{t}
 =\sqrt{\frac{\epsilon_t}{2\epsilon_c}}$,
can be written as,  
\begin{equation}
\gamma_{t}
 =\frac{1}{\sqrt{2}}\left(\frac{3^{3+\frac{s}{2}}}{2^{7}}
\sqrt{\frac{R}{\lambda}}
\frac{ l_{\rm nth}G_1 F(\Gamma_{\rm inj})}{l_{B}^{3/2}(1+g)}
\right)^{\frac{1}{4+s}}.
\label{eq:gtrad}
\end{equation}

If the particles emitting around the turn-over frequency are in the coulomb dominated regime ( i.e. the evaluation of equation~(\ref{eq:gtrad}) leads to  $\gamma_t < \sqrt{p_{\rm c}^2+1}$) then  Coulomb effects must be taken into account:
\begin{equation}
\epsilon_t=\left(\frac{3^{(3+s)}}{2^8}\frac{\pi}{\alpha_{\rm s}}\left(\frac{\tau_0}{\xi}\right)^2 G_1^2\right)^{1/(4+s)}\epsilon_c^{(2+s)/(4+s)},
\end{equation}
with $s$ now equal to the power-law index in the Coulomb regime $s=\Gamma_{\rm inj}-1$, and

\begin{equation}
\gamma_{t}
 =\frac{1}{\sqrt{2}}\left(\frac{3^{1+\frac{s}{2}}}{2^{4}}\sqrt{\frac{R}{\lambda l_{B}}}
\frac{\langle\gamma\rangle l_{\rm nth}}{\tau_{\rm T}\ln\Lambda}G_1 F(\Gamma_{\rm inj})  \right)^{\frac{1}{4+s}}.
\label{eq:gtcoul}
\end{equation}
According to both equations~(\ref{eq:gtrad}) an (\ref{eq:gtcoul}), $\gamma_{\rm t}$ is insensitive to the model parameters. In the regime studied in this work $\gamma_{\rm t}\simeq 10$. 

Now that we have an estimate of $\gamma_{\rm t}$ we can evaluate the synchrotron compactness (as defined by equation~(\ref{eq:lsydef})):
\begin{equation}
l_{\rm sync}\simeq\frac{16\pi}{9}l_{B}\int_{\gamma_{\rm t}}^{+\infty}\tau(\gamma)\gamma^2\beta^2d\gamma.
\label{eq:lsyncint}
\end{equation}

In the case where $\gamma_{\rm t}>\sqrt{p_{\rm c}^2+1}$ and in the relativistic limit, the synchrotron radiation compactness, controlling the Compton cooling of the Maxwellian component,  can be estimated as:
\begin{equation}
l_{sync}\simeq
\frac{l_{\rm nth}}{1+g}\frac{F(\Gamma_{\rm inj})}{\Gamma_{\rm inj}-2}
\gamma_{\rm t}^{2-\Gamma_{\rm inj}}.
\label{eq:lsync}
\end{equation}
Combining equation~(\ref{eq:lsync}) with (\ref{eq:g}),  and setting $\gamma_{\rm t}=10$, we can evalutate $g$ as:
\begin{eqnarray}
g&=&\sqrt{\left(\frac{3}{8\pi}\frac{l_{\rm s}}{l_B}(1+\tau_{\rm T}/3)+\frac{1}{2}\right)^{2}+\frac{3}{4\pi}\frac{l_{\rm nth}}{l_{B}}\frac{(1+\tau_{\rm T}/3)F(\Gamma_{\rm inj})}{(\Gamma_{\rm inj}-2)\gamma_{\rm t}^{-2+\Gamma_{\rm inj}}}}\nonumber\\
&&+\frac{3}{8\pi}\frac{l_{\rm s}}{l_B}(1+\tau_{\rm T}/3)-\frac{1}{2}.
\label{eq:geq}
\end{eqnarray}
Using this estimate of $g$, a  better approximation of $\gamma_{\rm t}$ can be obtained using equation~(\ref{eq:gtrad}), which in turn may be used  improve the estimate of $g$ and so on.  Since $\gamma_{\rm t}$ depends only weakly on $g$, convergence is usually achieved after a few iterations only.  
For  the general case a similar iterative procedure can be applied numerically using equation~(\ref{eq:lsyncint}).

The synchrotron boiler plays a role in the heating of the thermalised particles if, at  $\gamma_{\rm t}$, synchrotron dominates over Coulomb cooling:
\begin{equation}
\gamma_{\rm t}^2\beta_{\rm t}^3>\frac{9}{8}\frac{\ln\Lambda \tau_{\rm T}}{l_{\rm B} \langle\gamma\rangle}.
\label{eq:syncboilrules}
\end{equation}

Since $\gamma_{\rm t}\simeq10$, the synchrotron boiler is important  for thermal heating if  $l_{\rm B}/\tau_{\rm T} \ga1/5$. 

\bsp

\label{lastpage}


\begin{thebibliography}{99}

\bibitem[\protect\citeauthoryear{Abramowicz et 
al.}{1996}]{1996ApJ...471..762A} 
Abramowicz M.~A., Chen X.-M., Granath M., Lasota J.-P., 1996, ApJ, 471, 762 

\bibitem[\protect\citeauthoryear{Belmont, Malzac, 
\& Marcowith}{2008}]{2008MmSAI..79..228B} Belmont R., Malzac J., Marcowith A., 2008a, MmSAI, 79, 228,  arXiv:0802.2661

\bibitem[\protect\citeauthoryear{Belmont, Malzac, 
\& Marcowith}{2008}]{2008MmSAI..79..228B} Belmont R., Malzac J., Marcowith A., 2008b, A\&A, in press (arXiv:0808.1258). 

\bibitem[\protect\citeauthoryear{Beloborodov}{1999}]{1999ApJ...510L.123B} 
Beloborodov A.~M., 1999, ApJ, 510, L123 

\bibitem[\protect\citeauthoryear{Bisnovatyi-Kogan 
\& Lovelace}{1997}]{1997ApJ...486L..43B} 
Bisnovatyi-Kogan G.~S., Lovelace R.~V.~E., 1997, ApJ, 486, L43 

\bibitem[\protect\citeauthoryear{Blandford 
\& Begelman}{1999}]{1999MNRAS.303L...1B} Blandford R.~D., Begelman M.~C., 1999, MNRAS, 303, L1 

\bibitem[\protect\citeauthoryear{Bowyer et al.}{1965}]{1965Sci...147..394B} 
Bowyer S., Byram E.~T., Chubb T.~A., Friedman H., 1965, Sci, 147, 394 

\bibitem[\protect\citeauthoryear{Chaty et al.}{2003}]{2003MNRAS.346..689C} 
Chaty S., Haswell C.~A., Malzac J., Hynes R.~I., Shrader C.~R., Cui W., 
2003, MNRAS, 346, 689 

\bibitem[\protect\citeauthoryear{Coppi}{1992}]{1992MNRAS.258..657C} Coppi 
P.~S., 1992, MNRAS, 258, 657 

\bibitem[\protect\citeauthoryear{Coppi}{1999}]{1999ASPC..161..375C} Coppi 
P.~S., 1999, ASPC, 161, 375 

\bibitem[\protect\citeauthoryear{Del Santo et 
al.}{2008}]{2008MNRAS.390..227D} Del Santo M., Malzac J., Jourdain E., 
Belloni T., Ubertini P., 2008, MNRAS, 390, 227 

\bibitem[\protect\citeauthoryear{di Matteo, Blackman, 
\& Fabian}{1997}]{1997MNRAS.291L..23D} di Matteo T., Blackman E.~G., Fabian A.~C., 1997, MNRAS, 291, L23 

\bibitem[\protect\citeauthoryear{di Matteo, Celotti, 
\& Fabian}{1997}]{1997MNRAS.291..805D} di Matteo T., Celotti A., Fabian A.~C., 1997, MNRAS, 291, 805 

\bibitem[\protect\citeauthoryear{Done, Gierli{\'n}ski, 
\& Kubota}{2007}]{2007A&ARv..15....1D} Done C., Gierli{\'n}ski M., Kubota A., 2007, A\&ARv, 15, 1  (DGK07)

\bibitem[\protect\citeauthoryear{Dermer}{1986}]{1986ApJ...307...47D} Dermer 
C.~D., 1986, ApJ, 307, 47 


\bibitem[\protect\citeauthoryear{Esin, McClintock, 
\& Narayan}{1997}]{1997ApJ...489..865E} Esin A.~A., McClintock J.~E., Narayan R., 1997, ApJ, 489, 865 

\bibitem[\protect\citeauthoryear{Esin et al.}{1998}]{1998ApJ...505..854E} 
Esin A.~A., Narayan R., Cui W., Grove J.~E., Zhang S.-N., 1998, ApJ, 505, 
854 

\bibitem[\protect\citeauthoryear{Esin et al.}{2001}]{2001ApJ...555..483E} 
Esin A.~A., McClintock J.~E., Drake J.~J., Garcia M.~R., Haswell C.~A., 
Hynes R.~I., Muno M.~P., 2001, ApJ, 555, 483 


\bibitem[\protect\citeauthoryear{Fabian et al.}{1986}]{1986MNRAS.221..931F} 
Fabian A.~C., Guilbert P.~W., Blandford R.~D., Phinney E.~S., Cuellar L., 
1986, MNRAS, 221, 931 


\bibitem[\protect\citeauthoryear{Frontera et 
al.}{2001}]{2001ApJ...546.1027F} Frontera F., et al., 2001a, ApJ, 546, 1027 

\bibitem[\protect\citeauthoryear{Frontera et 
al.}{2001}]{2001ApJ...561.1006F} Frontera F., et al., 2001b, ApJ, 561, 1006 


\bibitem[\protect\citeauthoryear{Galeev, Rosner, 
\& Vaiana}{1979}]{1979ApJ...229..318G}
 Galeev A.~A., Rosner R., Vaiana G.~S., 1979, ApJ, 229, 318 

\bibitem[\protect\citeauthoryear{Ghisellini, Guilbert, 
\& Svensson}{1988}]{1988ApJ...334L...5G} Ghisellini G., Guilbert P.~W., Svensson R., 1988, ApJ, 334, L5 

\bibitem[\protect\citeauthoryear{Ghisellini, Haardt, 
\& Svensson}{1998}]{1998MNRAS.297..348G} Ghisellini G., Haardt F., Svensson R., 1998, MNRAS, 297, 348 



\bibitem[\protect\citeauthoryear{Gierli{\'n}ski et 
al.}{1999}]{1999MNRAS.309..496G} 
Gierli{\'n}ski M., Zdziarski A.~A., Poutanen J., Coppi P.~S., Ebisawa K., Johnson W.~N., 1999, MNRAS, 309, 496 (G99) 



\bibitem[\protect\citeauthoryear{Gilfanov, Churazov, 
\& Revnivtsev}{1999}]{1999A&A...352..182G} Gilfanov M., Churazov E., Revnivtsev M., 1999, A\&A, 352, 182 

\bibitem[\protect\citeauthoryear{Haardt, Maraschi, 
\& Ghisellini}{1994}]{1994ApJ...432L..95H} Haardt F., Maraschi L., Ghisellini G., 1994, ApJ, 432, L95 


\bibitem[\protect\citeauthoryear{Hirose, Krolik, 
\& Stone}{2006}]{2006ApJ...640..901H} Hirose S., Krolik J.~H., Stone J.~M., 2006, ApJ, 640, 901 



\bibitem[\protect\citeauthoryear{Ichimaru}{1977}]{1977ApJ...214..840I} 
Ichimaru S., 1977, ApJ, 214, 840 


\bibitem[\protect\citeauthoryear{Joinet et al.}{2007}]{2007ApJ...657..400J} 
Joinet A., Jourdain E., Malzac J., Roques J.~P., Corbel S., Rodriguez J., 
Kalemci E., 2007, ApJ, 657, 400 

\bibitem[\protect\citeauthoryear{Katarzy{\'n}ski et 
al.}{2006}]{2006A&A...453...47K} Katarzy{\'n}ski, K., Ghisellini, G., Mastichiadis, A., Tavecchio, F., \& Maraschi, L.\ 2006, A\&A, 453, 47

\bibitem[\protect\citeauthoryear{Li 
\& Miller}{1997}]{1997ApJ...478L..67L}
 Li H., Miller J.~A., 1997, ApJ, 478, L67 

\bibitem[\protect\citeauthoryear{Liang 
\& Price}{1977}]{1977ApJ...218..247L} 
Liang E.~P.~T., Price R.~H., 1977, ApJ, 218, 247 

\bibitem[\protect\citeauthoryear{Lightman 
\& Zdziarski}{1987}]{1987ApJ...319..643L} 
Lightman A.~P., Zdziarski A.~A., 1987, ApJ, 319, 643 

\bibitem[\protect\citeauthoryear{Magdziarz 
\& Zdziarski}{1995}]{1995MNRAS.273..837M}
 Magdziarz P., Zdziarski A.~A., 1995, MNRAS, 273, 837 

\bibitem[\protect\citeauthoryear{Mahadevan 
\& Quataert}{1997}]{1997ApJ...490..605M} Mahadevan R., Quataert E., 1997, ApJ, 490, 605 

\bibitem[\protect\citeauthoryear{Malzac}{2007}]{2007Ap&SS.311..149M} Malzac J., 2007, Ap\&SS, 311, 149 


\bibitem[\protect\citeauthoryear{Malzac, Beloborodov, 
\& Poutanen}{2001}]{2001MNRAS.326..417M} 
Malzac J., Beloborodov A.~M., Poutanen J., 2001, MNRAS, 326, 417 


\bibitem[\protect\citeauthoryear{Massey, Johnson, 
\& Degioia-Eastwood}{1995}]{1995ApJ...454..151M} Massey P., Johnson K.~E., Degioia-Eastwood K., 1995, ApJ, 454, 151 

\bibitem[\protect\citeauthoryear{McClintock et 
al.}{2001}]{2001ApJ...551L.147M} McClintock J.~E., Garcia M.~R., Caldwell 
N., Falco E.~E., Garnavich P.~M., Zhao P., 2001, ApJ, 551, L147 

\bibitem[\protect\citeauthoryear{McConnell et 
al.}{2002}]{2002ApJ...572..984M} 
McConnell M.~L., et al., 2002, ApJ, 572, 984 (MC02)

\bibitem[\protect\citeauthoryear{Merloni 
\& Fabian}{2001}]{2001MNRAS.321..549M} Merloni A., Fabian A.~C., 2001, MNRAS, 321, 549 

\bibitem[\protect\citeauthoryear{Miller 
\& Stone}{2000}]{2000ApJ...534..398M} Miller K.~A., Stone J.~M., 2000, ApJ, 534, 398 

\bibitem[\protect\citeauthoryear{Mitsuda et 
al.}{1984}]{1984PASJ...36..741M} 
Mitsuda K., et al., 1984, PASJ, 36, 741 

\bibitem[\protect\citeauthoryear{Narayan 
\& Yi}{1994}]{1994ApJ...428L..13N} Narayan R., Yi I., 1994, ApJ, 428, L13 

\bibitem[\protect\citeauthoryear{Nayakshin 
\& Melia}{1998}]{1998ApJS..114..269N}
 Nayakshin S., Melia F., 1998, ApJS, 114, 269 

\bibitem[\protect\citeauthoryear{Poutanen, Krolik, 
\& Ryde}{1997}]{1997MNRAS.292L..21P}
Poutanen J., Krolik J.~H., Ryde F., 1997, MNRAS, 292, L21 

\bibitem[\protect\citeauthoryear{Quataert 
\& Gruzinov}{1999}]{1999ApJ...520..248Q} Quataert E., Gruzinov A., 1999, ApJ, 520, 248 

\bibitem[\protect\citeauthoryear{R{\'o}{\.z}a{\'n}ska 
\& Czerny}{2000}]{2000A&A...360.1170R} R{\'o}{\.z}a{\'n}ska A., Czerny B., 2000, A\&A, 360, 1170 
 

\bibitem[\protect\citeauthoryear{Shakura 
\& Syunyaev}{1973}]{1973A&A....24..337S}
 Shakura N.~I., Syunyaev R.~A., 1973, A\&A, 24, 337 

\bibitem[\protect\citeauthoryear{Shapiro, Lightman, 
\& Eardley}{1976}]{1976ApJ...204..187S} Shapiro S.~L., Lightman A.~P., Eardley D.~M., 1976, ApJ, 204, 187 

\bibitem[\protect\citeauthoryear{Svensson}{1984}]{1984MNRAS.209..175S} 
Svensson R., 1984, MNRAS, 209, 175 

\bibitem[\protect\citeauthoryear{Sunyaev 
\& Titarchuk}{1980}]{1980A&A....86..121S}
 Sunyaev R.~A., Titarchuk L.~G., 1980, A\&A, 86, 121 

%\bibitem[\protect\citeauthoryear{Svensson}{1987}]{1987MNRAS.227..403S} 
%Svensson R., 1987, MNRAS, 227, 403 

\bibitem[\protect\citeauthoryear{Vurm 
\& Poutanen}{2008}]{2008arXiv0802.3680V} Vurm I., Poutanen J., 2008, Proceedings of the workshop 'High Energy Phenomena in Relativistic Outflows' (HEPRO), Dublin, Ireland, arXiv:0802.3680 

\bibitem[\protect\citeauthoryear{Wardzi{\'n}ski 
\& Zdziarski}{2000}]{2000MNRAS.314..183W} Wardzi{\'n}ski G., Zdziarski A.~A., 2000, MNRAS, 314, 183 

\bibitem[\protect\citeauthoryear{Wardzi{\'n}ski 
\& Zdziarski}{2001}]{2001MNRAS.325..963W} Wardzi{\'n}ski G., Zdziarski A.~A., 2001, MNRAS, 325, 963 , (WZ01)

\bibitem[\protect\citeauthoryear{Wardzi{\'n}ski et 
al.}{2002}]{2002MNRAS.337..829W} Wardzi{\'n}ski G., Zdziarski A.~A., 
Gierli{\'n}ski M., Grove J.~E., Jahoda K., Johnson W.~N., 2002, MNRAS, 337, 
829 

\bibitem[\protect\citeauthoryear{Yuan, Quataert, 
\& Narayan}{2003}]{2003ApJ...598..301Y} Yuan F., Quataert E., Narayan R., 2003, ApJ, 598, 301 

\bibitem[\protect\citeauthoryear{Yuan 
\& Zdziarski}{2004}]{2004MNRAS.354..953Y} Yuan F., Zdziarski A.~A., 2004, MNRAS, 354, 953 

\bibitem[\protect\citeauthoryear{Yuan, Cui, 
\& Narayan}{2005}]{2005ApJ...620..905Y} Yuan F., Cui W., Narayan R., 2005, ApJ, 620, 905 

\bibitem[\protect\citeauthoryear{Yuan et al.}{2007}]{2007ApJ...659..541Y} 
Yuan F., Zdziarski A.~A., Xue Y., Wu X.-B., 2007, ApJ, 659, 541 

\bibitem[\protect\citeauthoryear{Zdziarski, Lubi{\'n}ski, 
\& Smith}{1999}]{1999MNRAS.303L..11Z} Zdziarski A.~A., Lubi{\'n}ski P., Smith D.~A., 1999, MNRAS, 303, L11 

\bibitem[\protect\citeauthoryear{Zdziarski et 
al.}{2002}]{2002ApJ...578..357Z} Zdziarski A.~A., Poutanen J., Paciesas 
W.~S., Wen L., 2002, ApJ, 578, 357 

\bibitem[\protect\citeauthoryear{Zi{\'o}{\l}kowski}{2005}]{2005MNRAS.358..851Z} Zi{\'o}{\l}kowski J., 2005, MNRAS, 358, 851 

\end{thebibliography}
\end{document}